\documentclass{aastex631}
\usepackage{color}
\usepackage{amsmath}

\shorttitle{Speckle characterization}
\shortauthors{Hirata \& Merchant}

\graphicspath{{./}{figures/}}

\begin{document}

\title{Pixel centroid characterization with laser speckle and application to the {\slshape Nancy Grace Roman Space Telescope} detector arrays}

\author[0000-0002-2951-4932]{Christopher M. Hirata}
\affiliation{Center for Cosmology and Astroparticle Physics,
191 West Woodruff Avenue,
Columbus, Ohio 43210, USA}
\affiliation{Department of Physics,
191 West Woodruff Avenue,
Columbus, Ohio 43210, USA}
\affiliation{Department of Astronomy,
140 West 18th Avenue,
Columbus, Ohio 43210, USA}

\author{Christopher Merchant}
\affiliation{NASA Goddard Space Flight Center, Detector Systems Branch, 8800 Greenbelt Rd, Greenbelt, Maryland 20771, USA}


\begin{abstract}
The {\slshape Nancy Grace Roman Space Telescope} will use its wide-field instrument to carry out a suite of sky surveys in the near infrared. Several of the science objectives of these surveys, such as the measurement of the growth of cosmic structure using weak gravitational lensing, require exquisite control of instrument-related distortions of the images of astronomical objects. {\slshape Roman} will fly new large-format (4k$\times$4k) Teledyne H4RG-10 infrared detector arrays. This paper investigates whether the pixel centroids are located on a regular grid by projecting laser speckle patterns through a double slit aperture onto a non-flight detector array. We develop a method to reconstruct the pixel centroid offsets from the stochastic speckle pattern. Due to the orientation of the test setup, only $x$-offsets are measured here. We test the method both on simulations, and by injecting artificial offsets into the real images. We use cross-correlations of the reconstructions from different speckle realizations to determine how much of the variance in the pixel offset maps is signal (fixed to the detector) and how much is noise. After performing this reconstruction on $64\times 64$ pixel patches, and fitting out the best-fit linear mapping from pixel index to position, we find that there are residual centroid offsets in the $x$ (column) direction from a regular grid of 0.0107 pixels RMS (excluding shifts of an entire row relative to another, which our speckle patterns cannot constrain). This decreases to 0.0097 pix RMS if we consider residuals from a quadratic rather than linear mapping. These RMS offsets include both the physical pixel offsets, as well as any apparent offsets due to cross-talk and remaining systematic errors in the reconstruction. We comment on the advantages and disadvantages of speckle scene measurements as a tool for characterizing the pixel-level behavior in astronomical detectors.
\end{abstract}


\section{Introduction} \label{sec:intro}

One of the key results in modern observational cosmology has been the discovery that the expansion of the Universe is accelerating \citep{1998AJ....116.1009R, 1999ApJ...517..565P}. The characterization of this acceleration -- and more generally, tests of the cosmological model on the largest accessible scales -- has motivated a wide range of observational programs. These include measurement of the luminosity distance-redshift relation with Type Ia supernovae; statistics of the distortions of galaxy shapes by weak gravitational lensing; and clustering statistics in photometric and spectroscopic galaxy surveys (see, e.g., \citealt{2013PhR...530...87W} for a review). As larger focal planes have been deployed on modern wide-field telescopes, the precision of measurements of both the expansion history of the Universe and the growth of cosmic structure has improved. In the area of weak lensing, for example, we have proceeded from the early detections 20 years ago \citep{2000A&A...358...30V, 2000MNRAS.318..625B, 2000Natur.405..143W, 2000astro.ph..3338K} to recent measurements that reach few percent precision or better \citep{2020PASJ...72...16H, 2020A&A...634A.127A, 2022PhRvD.105b3514A}.

The weak lensing signal is very small ($\sim 1\%$ distortion on a typical line of sight), and we aim to measure it to high precision, so even small sources of systematic error -- whether astrophysical, observational, or introduced in data processing -- must be taken into account \citep[e.g.][]{2018ARA&A..56..393M}. The large next-generation weak lensing surveys -- including the Legacy Survey of Space and Time on the Vera Rubin Observatory \citep{2018arXiv180901669T}, and the weak lensing key projects on the {\slshape Euclid} telescope \citep{2011arXiv1110.3193L} and the {\slshape Nancy Grace Roman Space Telescope} \citep{2015arXiv150303757S} -- have thus had major efforts to characterize their instruments and incorporate non-ideal effects in simulations. This includes effects at the pixel level in the detectors that turn incoming electromagnetic waves into an electrical signal that is ultimately digitized and turned into a ``raw'' data file.

The {\slshape Roman Space Telescope} will fly a mosaic of the new Teledyne H4RG-10 4k$\times$4k near infrared (NIR) detector arrays using 2.5 $\mu$m cutoff Hg$_{1-x}$Cd$_x$Te as the light-sensitive material \citep{2020JATIS...6d6001M}. Access to the observer-frame near infrared is key for photometric redshifts \citep[e.g.][]{2019ApJ...877..117H} and supplies more source photons per unit area than are available in the visible for most galaxy spectra; we can make full use of these advantages with the low sky background in space, diffraction from a 2.4 m aperture, and the availability of large-format NIR arrays. However, these arrays are new to weak lensing, and the H4RG-10 specifically uses a smaller physical pixel size than the H2RG (2k$\times$2k) arrays that have become standard in ground- and space-based infrared astronomy. We have therefore made an extensive effort to characterize pixel-level effects in the {\slshape Roman} detectors. This has included past studies on the brighter-fatter effect, inter-pixel capacitance, non-linearity, and charge diffusion \citep{2020PASP..132a4501H, 2020PASP..132a4502C, 2020PASP..132g4504F, 2022PASP..134a4001G}.

One detector-related imprint that is of potential interest to {\slshape Roman} is displacement of the centroids from a regular grid. If present, such displacements could affect galaxy shapes either directly (since the galaxy surface brightness is not being sampled at the position of the regular grid), or through point spread function (PSF) determination since the bright stars are not being sampled on a regular grid, either. Pixel centroid offsets could also be important for other science programs on {\slshape Roman}, such as astrometric shifts in microlensing events \citep[e.g.][]{2014ApJ...784...64G, 2016ApJ...830...41L, 2022arXiv220201903L}, proper motions and parallaxes \citep[e.g.][]{ 2019BAAS...51c.211G, 2019JATIS...5d4005W}. Point source photometry (used for both the microlensing and the supernova program; \citealt{2018ApJ...867...23H}) may also be affected since the {\slshape Roman} PSF is of order one pixel wide. Some even more demanding astrometric applications have also been proposed \citep[e.g.][]{2015JKAS...48...93G, 2018AJ....155..102M, 2021MNRAS.501.2688C}. Pixel centroid shifts are well-motivated from studies of previous generations of NIR detectors. Examples of features reported in H2RGs include sub-pixel sensitivity variations associated with the crystal planes of the Hg$_{1-x}$Cd$_x$Te (see, e.g., studies on non-flight {\slshape Euclid} detectors; \citealt{2018SPIE10709E..36S}) and defects unique to individual pixels \citep{2007PASP..119..466B, 2014SPIE.9154E..2DH}. Apparent pixel centroid shifts due to settling effects (which can bias astrometry even if the physical pixels are on a regular grid) have also been reported on ground-based instruments \citep{2014A&A...563A..80L}, and cross-talk effects that could produce a similar but much smaller effect have been observed during {\slshape Roman} development \citep{2020PASP..132g4504F}.

Pixel centroid offsets cannot be measured from flat field illumination, as they can only be revealed by observing scenes with spatial structure. Astronomical observations can provide a very detailed test of pixel offsets, and such a test has been included in the {\slshape Roman} on-orbit calibration plan, but of course the data will not be available until after launch, and so it is strongly desirable to have laboratory tests to guide planning and pipeline development. The {\slshape Roman} project is planning a set of more detailed characterization tests that will project a deterministic pattern onto detector arrays that were built in the flight lots but were not selected for flight.

The investigation described in this paper was aimed at using the earliest available characterization data to get a ``first look'' at the pixel offset characteristics of an H4RG-10 detector array from the {\slshape Roman} flight lot. The data available when this investigation began came from a laser speckle projector that was used to measure the modulation transfer function (MTF). Speckle patterns are stochastic, and we had to develop appropriate statistical techniques to extract pixel offsets from observations of the speckle pattern. The purpose of this paper is two-fold: first, to present what we have learned about the pixel offsets in one of the {\slshape Roman} arrays; and second, to document the methodology as it may be useful in future projects if the available test capabilities support speckle projection more easily than a deterministic pattern.

This paper is organized as follows. In Section~\ref{sec:description}, we describe the available data and the basic properties of speckle fringes. We introduce the principles behind the pixel centroid analysis in Section~\ref{sec:concept}, and the mathematical formalism in Section~\ref{sec:formalism}. Our reconstruction algorithm is described in Section~\ref{sec:5}. We apply the method to simulations in Section~\ref{sec:sims}, and to the data on one of the H4RG-10 arrays in Section~\ref{sec:application}. We investigate the impact of the pixel offsets on {\slshape Roman} PSF fitting in Section~\ref{sec:psf}. We conclude in Section~\ref{sec:discussion}.

\section{Description of the available data}
\label{sec:description}

The data available for this analysis were taken in the Detector Characterization Laboratory at NASA Goddard Space Flight Center in January 2021. A detector array from the flight lot but that was not selected for flight (SCA21536) was used for this test.

\begin{figure}
\centering{\includegraphics[width=6.5in]{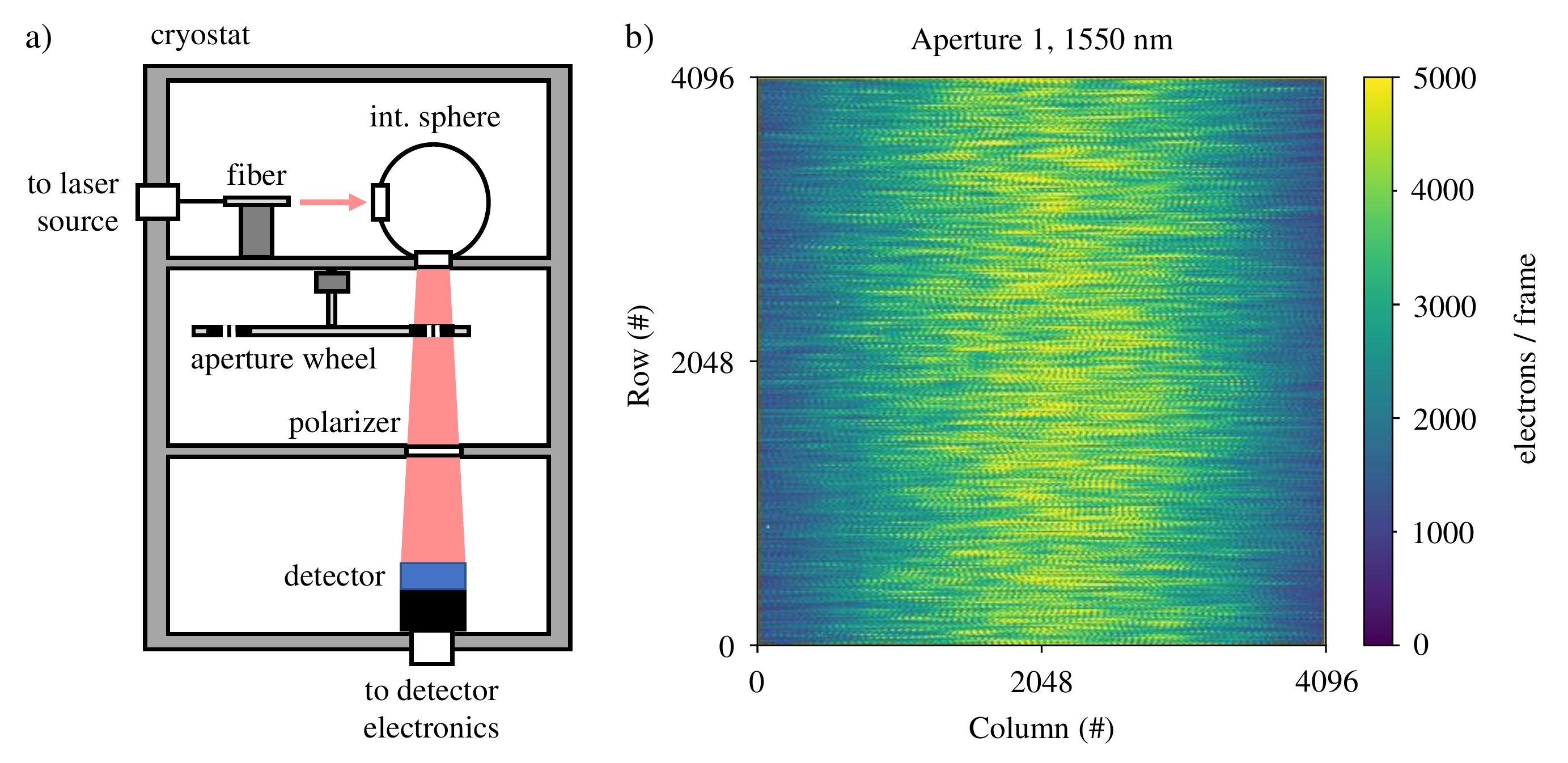}}
\caption{\label{fig:merchant}(a) A diagram of the test setup; and (b) an example CDS image of a fringe pattern using the AP1 aperture at 1.55 $\mu$m.}
\end{figure}

A diagram of the test setup is shown in Figure~\ref{fig:merchant}(a). The design is based on previous narrowband speckle generators \citep{1993OptEn..32..395S, 2005ApOpt..44.1543P}, except that the hardware is contained within a low-temperature cryostat. Enclosing the hardware within the cryostat permits simplified testing on infrared detectors. Laser speckle is generated within the setup by directing laser light from a fiber into an integrating sphere. The speckle emerging from the sphere illuminates a double-slit aperture, projecting randomly-positioned interference patterns onto the surface of the detector. A linear polarizer selects only one polarization of light (this increases the contrast $\delta I/\bar I$ of the fringes; it is not strictly necessary for the application described here). An example correlated double sample (CDS) image, captured by the detector, is shown in Fig.~\ref{fig:merchant}(b).

The spatial frequencies of the interference patterns, aligned along the detector rows, are determined by the distance $z_0$ from the aperture to the detector, the wavelength $\lambda$ of illumination, and the spacing $d_{\rm slit}$ between the slits in the aperture.
Every point in the aperture has a random emerging electric field, which can be considered a source illuminating the detector array in accordance with Huygens's principle. Every pair of points separated by a distance $(\Delta X,\Delta Y)$ in the aperture produces a fringe on the detector with wave vector ${\boldsymbol u} = (\Delta X,\Delta Y)P/(\lambda z_0)$, where ${\boldsymbol u}$ is in cycles per pixel, and $P = 10$ $\mu$m is the pixel pitch.  The result is a pattern of fringes on the detector whose 2D power spectrum is the autocorrelation of the aperture (rescaled by $P/\lambda z_0$).

\begin{figure*}
\centering{\includegraphics[width=5.5in]{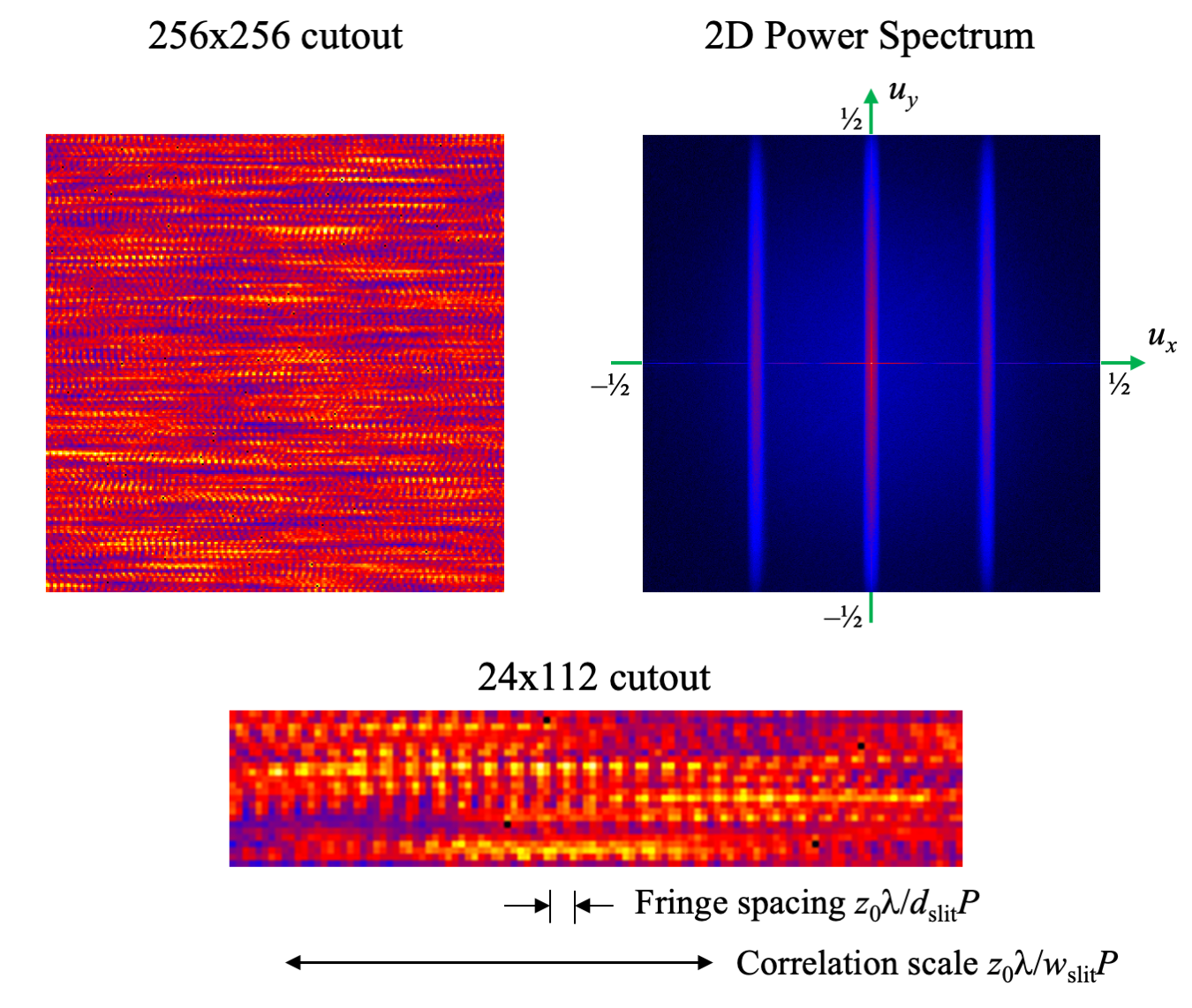}}
\caption{\label{fig:1}Examples of speckle fringe data obtained with SCA21536 and aperture AP7. {\em Left}: A $256\times 256$ cutout from the center of the image. {\em Right}: The 2D power spectrum of the image, displayed over the first Brillouin zone (Fourier wave vector components $u_x$ and $u_y$ between $-\frac12$ and $+\frac12$ cycles per pixel). {\em Bottom}: An enlarged cutout showing the individual fringes. The fringe spacing is $\lambda z_0/(d_{\rm slit}P) \approx 3.9$ pixels, and the fringes have a correlation length in the horizontal direction of $\lambda z_0/(w_{\rm slit}P)\approx 66$ pixels.}
\end{figure*}

The apertures used in the current setup are double slits of separation $d_{\rm slit}$ and width $w_{\rm slit}$. Since the autocorrelation of two slits is three vertical bars, the 2D power spectrum of the intensity pattern consists of three vertical bars at $u_x = 0$ and $u_x = \pm d_{\rm slit}P/(\lambda z_0)$ each with half-width $\Delta u_x = w_{\rm slit}P/(\lambda z_0)$ (see Fig.~\ref{fig:1}, right panel). In real space, this results in a fringe pattern with repeating maxima separated in the $x$ direction by $\lambda z_0/(d_{\rm slit}P)$ pixels (Fig.~\ref{fig:1}, left panel). If one moves in the horizontal direction, the amplitude and phase of the fringe pattern changes; it completely de-correlates after one moves a distance of $1/\Delta u_x = \lambda z_0/(w_{\rm slit} P)$ pixels (Fig.~\ref{fig:1}, bottom panel). If one moves in the vertical direction, the fringe pattern should de-correlate after one moves a distance of $\lambda z_0/(\ell_{\rm slit} P)$ pixels, where $\ell_{\rm slit}$ is the height of the slit. Since the slits are much taller than they are wide, this vertical de-correlation length is very short.

Small path length differences lead to reduced spatial frequencies (increased fringe spacing) moving away from the centerline of the detector. If one moves along the vertical direction, the reduction in spatial frequency follows a $\sim \cos\theta$ law (due to the aperture-detector image being $z_0/\cos\theta$ instead of $z_0$), whereas in the horizontal direction it follows the steeper $\sim \cos^3\theta$ law (one additional factor each of $\cos\theta$ due to the foreshortening of both the double-slit aperture separation $d_{\rm slit}$ and the pixel pitch $P$). This effect is reproduced in simulations based on the Fresnel diffraction integral (\S\ref{ss:fresnel}).

By adjusting some aspect of the system -- in this case, the position of the fiber that feeds the laser light into the integrating sphere -- one may obtain a new realization of the speckle pattern. The data set used here has a sequence of apertures (AP$n$ for $n = 1...12$, each with increasing slit separation $d_{\rm slit}$), and 25 realizations of the speckle pattern for each aperture. The wavelength of light used is $\lambda = 1.55$ $\mu$m. The last two apertures (AP11 and AP12) have fringe spacings that go past the Nyquist frequency, $d_{\rm slit}P/(\lambda z_0)\ge\frac12$. The images obtained have 5 up-the-ramp samples; at 200 kHz readout with 32 channels, and including overheads, the spacing of each sample is 2.83 s.

The apertures AP4 through AP9 were used in this analysis (to give a range of spatial frequencies, but to stay below the Nyquist frequency). The parameters of these apertures are given in Table~\ref{tab:apertures}.

\begin{table}[]
    \centering
    \begin{tabular}{c|cc}
\hline
        Aperture & Slit spacing & Slit width \\
         & $d_{\rm slit}$ [mm] & $w_{\rm slit}$ [mm] \\ \hline
         AP4 & 2.0 & 0.2 \\
         AP5 & 3.0 & 0.2 \\
         AP6 & 4.0 & 0.2 \\
         AP7 & 5.0 & 0.3 \\
         AP8 & 6.0 & 0.3 \\
         AP9 & 7.0 & 0.3 \\
\hline
    \end{tabular}
    \caption{\label{tab:apertures}Apertures used in this analysis.}
\end{table}

The detector arrays were operated at a temperature of 95 K and bias voltage of 1.0 V for this test. A Leach controller rather than a flight-like controller was used; however for pixel behaviors such as centroid offsets that are in the Sensor Chip Assembly, the laboratory controller is expected to provide useful data. It is possible that apparent astrometric shifts that are related to settling effects will be different in the flight configuration.

\section{The concept}
\label{sec:concept}

The speckle data were originally acquired to measure the MTF of the {\slshape Roman} detector arrays. This can be obtained from the 2D power spectrum of the images  \citep{1993OptEn..32..395S, 2005ApOpt..44.1543P}. However, one might also expect that the speckle fringe data could be useful for measuring the astrometric offset of individual pixels. If one thinks about this in {\em real space}, then the fringe pattern should allow us to determine if a pixel is offset to the right or left relative to the grid pattern established by the other pixels in that row. This would not enable any kind of absolute measurement of that pixel's position -- distortions of the grid of pixels that are on scales larger than $\lambda z_0/(w_{\rm slit} P)$ would be lost. Additionally, distortions that move an entire row left or right relative to other rows would also be lost. If one thinks about this in {\em Fourier space}, then if the pixels are not on a regular grid but have their own individual astrometric offsets, this will take a pure Fourier mode $e^{2\pi i{\boldsymbol u}\cdot{\boldsymbol r}}$ in the speckle pattern and produce power in all of the other Fourier modes. Specifically, if the distortion map contains Fourier mode ${\boldsymbol u}'$, then the observed map will contain Fourier modes ${\boldsymbol u}\pm{\boldsymbol u}'$. The double slit speckle pattern is characterized by very high power in some regions of the ${\boldsymbol u}$-plane, and very low power in others. So by looking for the modes at ${\boldsymbol u}\pm{\boldsymbol u}'$ that should not be present for an ideal detector, we may constrain the positional offsets of the individual pixels.

Putting these ideas into practice is a challenge because speckle patterns are a random field: no finite region of the detector has a speckle pattern with fixed amplitude and phase. This can be contrasted with, e.g., two-beam laser interference patterns that are sinusoidal over an entire reconstruction patch \citep[e.g.][]{1995ApOpt..34.6672S, 2011SPIE.8151E..0WN, 2016A&A...595A.108C}. Fortunately, there is a long history of studying distortions of random fields in the context of lensing of the cosmic microwave background (CMB) radiation. The temperature and polarization fluctuations of the CMB are a random field, and they are distorted by gravitational lensing from intervening large scale structure. In CMB lensing reconstruction, one aims to learn about the lensing distortion field; one must do so without knowledge of the particular realization of the CMB anisotropy pattern that the Universe presents (since only the lensed field is observable). The most powerful way to measure the lensing field from the CMB data is not to use the changes in the power spectrum, but to use the correlations among the Fourier modes of the CMB induced by lensing \citep{1997A&A...324...15B, 2000PhRvD..62d3007H, 2001ApJ...557L..79H, 2002ApJ...574..566H}. These techniques led to the first detections of lensing of the CMB \citep{2007PhRvD..76d3510S, 2008PhRvD..78d3520H} and are now part of the standard toolbox to extract cosmological information from the CMB \citep[e.g.][]{2020A&A...641A...8P}. This toolbox can be copied over to the problem of measuring astrometric offsets of pixels in observations of a speckle pattern. Mathematically, the primordial CMB anisotropies are analogous to the optical speckle pattern, and the lensing field in the CMB case is analogous to the field of pixel astrometric offsets.

There are some differences between CMB lensing reconstruction and pixel offset reconstruction, however none of these are fundamental:
\begin{itemize}
    \item The CMB is on a curved sky (the celestial sphere; \citealt{2003PhRvD..67h3002O}), whereas a {\slshape Roman} detector array is a flat square with boundaries. (This means that we really work with Fourier modes instead of spherical harmonics.)
    \item The CMB is, to a very good approximation, a Gaussian random field \citep{2020A&A...641A...9P}. Any small patch of the speckle pattern should have an electric field that is a complex Gaussian field because we can apply the central limit theorem to the superposition of reflection paths through the integrating sphere to the double slit aperture. However, the observable intensity is the square of the electric field and thus is not Gaussian. This does not represent a challenge since we build our estimators out of quadratic combinations of the data.
    \item There is only one realization of the CMB in the observable Universe (leading to cosmic variance limitations), whereas we can generate many realizations of the speckle fringe pattern in the laboratory and observe them with the same fixed pattern of detector peculiarities. This also allows us to cross-correlate reconstructions from different realizations, which eliminate the need for the Gaussianity assumption in the CMB lensing formalism (see discussion in \citealt{2006ApJ...653..922Z} and \citealt{2008MNRAS.388.1819L} in the context of multiple source planes in lensing of the cosmological 21 cm radiation).
    \item The CMB lensing field is -- to a good approximation -- described by the gradient of a scalar potential since lensing deflection comes from variations in gravitational potential. However, pixel offsets in a detector are in principle 2 independent functions of position ($x$ and $y$ offset at each pixel), and there is no reason to think that they would be the gradient of a scalar. This is not fundamental challenge since the CMB lensing formalism has been extended to 2 deflection components \citep{2005PhRvD..71l3527C, 2012JCAP...01..007N}.
    \item {\slshape Roman} detectors are not polarization sensitive\footnote{Small polarization effects may be possible due to the anti-reflection coating \citep{2020JATIS...6d6001M}, or to physical optics effects when the wavelength of light becomes comparable to the pixel size (\citealt{2010JOSAA..27.1219T}; here the pixel size is $P = 6.45\lambda$).}, so unlike the case of the CMB the Stokes $Q$ and $U$ parameters are not observable.
\end{itemize}

While most of the CMB theory literature works in Fourier (or spherical harmonic) space, the covariance matrices and associated linear algebra operations for CMB lensing reconstruction in observational work are usually written in real space. This enables simple handling of analysis regions that are a subset of the celestial sphere, for example due to the survey boundaries, Galactic plane cuts, or masking of microwave point sources. We make the same choice here, since our detector has boundaries (and we usually investigate sub-regions), and there are inoperable pixels that need to be masked.

We next proceed to describe our implementation of the CMB lensing estimators in the context of speckle illumination of {\slshape Roman} detector arrays.


\section{Formalism: Pixels and response functions}
\label{sec:formalism}

We consider a grid of $N$ pixels that -- in an ideal detector -- would be located at integer positions on the 1 and 2 axes. In principle, this might be all $4088^2$ pixels in a {\slshape Roman} detector, but in our application $N$ will be a smaller group of pixels such as $64^2$. If a pixel $i$ has a nominal position ${\boldsymbol r}_i\in{\mathbb Z}^2$, and there is an illumination pattern $I({\boldsymbol r})$, then we suppose that the pixel has a response function $R_i$ and that the signal observed in the detector is
\begin{equation}
S_i = {\cal A} F_i \int_{{\mathbb R}^2} I({\boldsymbol r})\,R_i({\boldsymbol r} - {\boldsymbol r}_i)\,d^2{\boldsymbol r},
\label{eq:Si}
\end{equation}
where ${\cal A}$ is some normalization constant and $F_i$ is the flat field of pixel $i$. In this equation, position is in units of the ideal pixel pitch (10 $\mu$m for {\slshape Roman} detectors). If we write functions in Fourier space via the transformation convention
\begin{equation}
\tilde f({\boldsymbol u}) = \int_{{\mathbb R}^2} f({\boldsymbol r})\,e^{-2\pi i {\boldsymbol u}\cdot{\boldsymbol r}}\,d^2{\boldsymbol r}
~~~\leftrightarrow~~~
f({\boldsymbol r}) = \int_{{\mathbb R}^2} \tilde f({\boldsymbol u})\,e^{2\pi i {\boldsymbol u}\cdot{\boldsymbol r}}\,d^2{\boldsymbol u},
\end{equation}
then we have
\begin{equation}
S_i = {\cal A}F_i \int_{{\mathbb R}^2}
e^{2\pi i{\boldsymbol u}\cdot{\boldsymbol r}_i}
\tilde I({\boldsymbol u}) \tilde R_i({\boldsymbol u})
\,d^2{\boldsymbol u}.
\end{equation}

Now we suppose that the response functions can be written as a sum over basis functions:
\begin{equation}
R_i({\boldsymbol r}) = \bar R({\boldsymbol r}) + \sum_{\alpha = 0}^{K-1} \xi_{i\alpha} B_\alpha({\boldsymbol r})
~~~\rightarrow ~~~
\tilde R_i({\boldsymbol u}) = \tilde{\bar R}({\boldsymbol u}) + \sum_{\alpha = 0}^{K-1} \xi_{i\alpha} \tilde B_\alpha({\boldsymbol u}).
\label{eq:RBasis}
\end{equation}
Here $i = 0...N-1$ is a pixel index\footnote{Since coding on this project is in Python, we use indices that start at 0.}; $\bar R({\boldsymbol r})$ is some reference response function (which could in principle be an estimate of the mean response function of a group of pixels); and $\alpha = 0...K-1$ describes which basis function is under consideration. The code in principle enables the use of up to $K=6$ of these basis functions, describing the normalization, centroid shift, and second moment shift of each pixel. The functions, and the interpretation of their coefficients, are:
\begin{equation}
\begin{array}{lcl}
\tilde B_0({\boldsymbol u}) = \tilde{\bar R}({\boldsymbol u}) & & {\rm flat~field~error},
\\
\tilde B_1({\boldsymbol u}) = -iu_x\tilde{\bar R}({\boldsymbol u}) & & x~{\rm astrometric~offset},
\\
\tilde B_2({\boldsymbol u}) = -iu_y\tilde{\bar R}({\boldsymbol u}) & & y~{\rm astrometric~offset},
\\
\tilde B_3({\boldsymbol u}) = -\frac12u_x^2\tilde{\bar R}({\boldsymbol u}) & & xx~{\rm second~moment},
\\
\tilde B_4({\boldsymbol u}) = -u_xu_y\tilde{\bar R}({\boldsymbol u}) & & xy~{\rm second~moment},
\\
\tilde B_5({\boldsymbol u}) = -\frac12u_y^2\tilde{\bar R}({\boldsymbol u}) & & yy~{\rm second~moment}.
\end{array}
\end{equation}
However, with the present set of data, we are attempting only to investigate the astrometric offsets.
The principal objective here is the determination of the $NK$ coefficients $\xi_{i\alpha}$. If some of these cannot be determined from the speckle data, then we would like to compute the ones that can be determined, and we would like to know of any degeneracies or directions in $NK$-dimensional space that cannot be reconstructed. Note that there will always be some combinations of coefficients that cannot be determined, for example, only the relative astrometric offsets of the pixels enter into observations of a random field (the ``common mode'' astrometric offset of all the pixels can only be determined if there is some information that identifies position $(0,0)$).

\section{Likelihood approach}
\label{sec:5}

We make a length $N$ data vector ${\boldsymbol d}$ out of the signals observed in $N$ pixels $i=0...N-1$. The data vector is obtained by taking the observed flat-fielded signal $S_i/F_i$; dividing an estimate of the mean $\overline{S/F}$ (see \S\ref{ss:prac} for the procedure we used); and subtracting 1:
\begin{equation}
d_i = \frac{S_i/F_i}{\overline{S/F}}-1.
\end{equation}
Thus the data vector is dimensionless and has mean zero. There is one such data vector for each realization of the speckle pattern for each aperture.

The data vector has a covariance matrix ${\bf C}$ that depends on the power spectrum of the illumination. There is a different power spectrum for each aperture, but each realization of the speckle pattern with the same aperture and wavelength is drawn from the same power spectrum and hence has the same covariance matrix ${\bf C}$. We suppose that the detector array is illuminated with a pattern that has an intensity power spectrum $P_I({\boldsymbol u})$. This power spectrum is defined by
\begin{equation}
\delta_I({\boldsymbol r}) = \frac{I({\boldsymbol r})}{\bar I} - 1~~~{\rm and} ~~~
\langle \tilde \delta_I^\ast({\boldsymbol u})
\tilde \delta_I({\boldsymbol u}') \rangle
= P_I({\boldsymbol u})\,\delta^{(2)}({\boldsymbol u}-{\boldsymbol u}').
\end{equation}
Then the covariance between two entries of the data vector $d_i$ and $d_j$ is
\begin{equation}
C_{ij} =\langle d_id_j\rangle = \int_{{\mathbb R}^2} P_I({\boldsymbol u}) \tilde R_i^\ast({\boldsymbol u})\tilde R_j({\boldsymbol u}) \, e^{2\pi i{\boldsymbol u}\cdot({\boldsymbol r}_i - {\boldsymbol r}_j)}\, d^2{\boldsymbol u} + {\cal N}_{ij},
\end{equation}
where ${\cal N}_{ij}$ is the noise covariance (including Poisson and read noise).
This can be written using Eq.~(\ref{eq:RBasis}):
\begin{eqnarray}
C_{ij} &=&~ \int_{{\mathbb R}^2} P_I({\boldsymbol u}) 
\left[\tilde {\bar R}^\ast({\boldsymbol u})\tilde {\bar R}({\boldsymbol u})
+ \sum_{\alpha=0}^{K-1} \xi_{i\alpha}
\tilde B_\alpha^\ast({\boldsymbol u})\tilde {\bar R}({\boldsymbol u})
+ \sum_{\alpha=0}^{K-1} \xi_{j\alpha}
\tilde {\bar R}^\ast({\boldsymbol u})\tilde B_\alpha({\boldsymbol u})
+ \sum_{\alpha=0}^{K-1}\sum_{\beta=0}^{K-1} \xi_{i\alpha}\xi_{j\beta}
\tilde B_\alpha^\ast({\boldsymbol u})\tilde B_\beta({\boldsymbol u})
\right]
\nonumber \\ && ~~~~\times e^{2\pi i{\boldsymbol u}\cdot({\boldsymbol r}_i - {\boldsymbol r}_j)}\, d^2{\boldsymbol u} + {\cal N}_{ij}.
\label{eq:Cij}
\end{eqnarray}
Thus we see that the covariance can be written as a quadratic function of the parameters $\xi_{i\alpha}$.

We now want to find the ``best fit'' values of the parameters $\xi_{i\alpha}$ for a given data set. We first write the Gaussian approximation to the log-likelihood function:
\begin{equation}
\ln{\cal L}_{\rm Gauss} = -\frac12{\boldsymbol d}^{\rm T}{\bf C}^{-1}{\boldsymbol d} -\frac12\ln\det{\bf C} + {\rm constant}.
\end{equation}
If there are $N_{\rm R}$ independent realizations of the data (in this case: $N_{\rm R}$ different fringe patterns obtained by dithering the position of the laser fiber), ${\boldsymbol d}^{(1)}...{\boldsymbol d}^{(N_{\rm R})}$, then we may sum the likelihoods. In practice, rather than taking all $N_{\rm R}$, we subtract the sample mean realization $\bar{\boldsymbol d} = \sum_{\mu=0}^{N_{\rm R}-1}{\boldsymbol d}^{(\mu)}/N_{\rm R}$. This results in only $N_{\rm R}-1$ independent realizations instead of $N_{\rm R}$, so the likelihood is
\begin{equation}
\ln{\cal L}_{\rm Gauss} = -\frac12 \sum_{\mu=0}^{N_{\rm R}-1}({\boldsymbol d}^{(\mu)}-\bar{\boldsymbol d})^{\rm T}{\bf C}^{-1}({\boldsymbol d}^{(\mu)}-\bar{\boldsymbol d}) -\frac12(N_{\rm R}-1)\ln\det{\bf C} + {\rm constant}.
\end{equation}
We can take the derivative of this with respect to a particular parameter:
\begin{equation}
\frac{\partial}{\partial\xi_{i\alpha}}\ln{\cal L}_{\rm Gauss} = 
\frac12 \sum_{\mu=0}^{N_{\rm R}-1}({\boldsymbol d}^{(\mu)}-\bar{\boldsymbol d})^{\rm T}{\bf C}^{-1} \frac{\partial{\bf C}}{\partial\xi_{i\alpha}}{\bf C}^{-1}({\boldsymbol d}^{(\mu)}-\bar{\boldsymbol d})
- \frac12N_{\rm R} {\rm Tr} \left[ {\bf C}^{-1}\frac{\partial{\bf C}}{\partial\xi_{i\alpha}} \right].
\label{eq:grad1}
\end{equation}
The maximum likelihood point is where Eq.~(\ref{eq:grad1}) is equal to zero for all $\xi_{i\alpha}$.

One potential issue with this procedure, which is different from the case of CMB lensing, is that the distribution of the data vector ${\boldsymbol d}$ is not Gaussian. This is a challenge for us if we want the true likelihood surface. However, what we are most interested in is that in the limit of lots of data, our maximum converges to the correct value. This should occur if the average value of Eq.~(\ref{eq:grad1}) is zero for the ``correct'' parameters. One can check that this condition is satisfied as long as $\langle d^{(\mu)}_i d^{(\mu)}_j\rangle = C_{ij}$, whether or not the distribution is Gaussian. In fact, we can go further and see that this condition is satisfied even if instead of the true ${\bf C}^{-1}$, we use an estimated $\hat{\bf C}^{-1}$:
\begin{equation}
\frac{\partial}{\partial\xi_{i\alpha}}\ln{\cal L}_{\rm Gauss} = 
\frac12 \sum_{\mu=0}^{N_{\rm R}-1}({\boldsymbol d}^{(\mu)}-\bar{\boldsymbol d})^{\rm T}\hat{\bf C}^{-1} \frac{\partial{\bf C}}{\partial\xi_{i\alpha}}\hat{\bf C}^{-1}({\boldsymbol d}^{(\mu)}-\bar{\boldsymbol d})
- \frac12(N_{\rm R}-1) {\rm Tr} \left[ \hat{\bf C}^{-1}{\bf C}\hat{\bf C}^{-1}\frac{\partial{\bf C}}{\partial\xi_{i\alpha}} \right].
\label{eq:grad2}
\end{equation}

In some cases, we may have parameters that are poorly determined. In this case, we write a prior on the parameters:
\begin{equation}
\Pi \propto \exp\left[ -\frac12 {\boldsymbol\xi}^{\rm T} {\boldsymbol\Sigma}^{-1}{\boldsymbol\xi}
\right] ~~~ \rightarrow ~~~
\frac{\partial}{\partial\xi_{i\alpha}}\ln\Pi = -[{\boldsymbol\Sigma}^{-1}{\boldsymbol\xi}]_{i\alpha},
\label{eq:prior}
\end{equation}
where we have written ${\boldsymbol\xi}$ as a length $NK$ vector and written an $NK\times NK$ prior covariance matrix ${\boldsymbol\Sigma}$.

The mathematical problem that we solve to find the maximum posterior probability point for the $\xi_{i\alpha}$ is then:
\begin{equation}
G_{i\alpha} \equiv
\sum_{\rm apertures} \left\{
\frac12 \sum_{\mu=0}^{N_{\rm R}-1}({\boldsymbol d}^{(\mu)}-\bar{\boldsymbol d})^{\rm T}\hat{\bf C}^{-1} \frac{\partial{\bf C}}{\partial\xi_{i\alpha}}\hat{\bf C}^{-1}({\boldsymbol d}^{(\mu)}-\bar{\boldsymbol d})
- \frac12(N_{\rm R}-1) {\rm Tr} \left[ \hat{\bf C}^{-1}{\bf C}\hat{\bf C}^{-1}\frac{\partial{\bf C}}{\partial\xi_{i\alpha}} \right]\right\}
-[{\boldsymbol\Sigma}^{-1}{\boldsymbol\xi}]_{i\alpha} = 0.
\label{eq:GP}
\end{equation}
This is a system of $NK$ equations for $NK$ unknowns (the $\xi_{i\alpha}$). It can be solved iteratively.

\subsection{Practicalities}
\label{ss:prac}

There are some important practical choices to make in Eq.~(\ref{eq:GP}).

First is the issue of bad pixels. Since Eq.~(\ref{eq:GP}) is working with the $N\times N$ matrix ${\bf C}$, it is straightforward to mask a pixel by striking the rows and columns corresponding to that pixel. A pixel may in principle be masked for only some of the apertures, or it could be masked in all apertures. In the latter case, the $\xi_{i\alpha}$ for a masked pixel $i$ only appears in the prior term in Eq.~(\ref{eq:GP}), and the prior will determine its estimated value.

Next is the choice of estimated $\hat{\bf C}^{-1}$. For this analysis, we compute the 2D power spectrum of the full detector array for each aperture (i.e., we do a fast Fourier transform [FFT] of the mean-subtracted $N_{\rm PS}\times N_{\rm PS}$ images, and average their squared amplitudes over all $N_{\rm R}$ realizations for that aperture). This produces a power spectrum on a $N_{\rm PS}\times N_{\rm PS}$ grid in ${\boldsymbol u}$, with samples spaced by $\Delta u_x = \Delta u_y = 1/N_{\rm PS}$ on each axis. In principle we would use the whole SCA ($N_{\rm PS}=4096$), however due to geometric distortion the power spectrum varies over the SCA. To test the effects of this, we repeat our analyses with both $N_{\rm PS}=1024$ and $N_{\rm PS}=512$. The power spectrum so obtained, $P_{\rm obs}({\boldsymbol u})$, is an estimate of
\begin{equation}
P_{\rm obs}({\boldsymbol u})\approx
P_I({\boldsymbol u})|\tilde{\bar R}({\boldsymbol u})|^2
\label{eq:Pobs}
\end{equation}
(that is, it includes the smearing from the pixel response function).
A few issues arise here. First, $P_{\rm obs}({\boldsymbol u})$ includes the noise (Poisson + read noise) as well as true fluctuations of the intensity. This does not represent a major issue for us, because the intensity fluctuations $\delta_I$ in speckle fringe patterns are of order unity, whereas Poisson fluctuations are of order 1\% (if there are typically $\sim 10^4$ elementary charges collected per pixel) and read noise is even less. A more serious issue is that $P_{\rm obs}({\boldsymbol u})$ as computed by the FFT is obtained over the first Brillouin zone,
\begin{equation}
{\cal B}_1 = \left\{ (u_x,u_y): ~
-\frac12<u_x<\frac12~{\rm and}~-\frac12<u_y<\frac12
\right\}.
\end{equation}
The current implementation of the code is limited to oversampled data, i.e., where the speckle pattern only contains Fourier modes in ${\cal B}_1$. This means that AP9 was the widest slit separation incorporated in this analysis. (The method could be easily extended to undersampled data in cases where no two Fourier modes present in the speckle pattern alias to each other. If two or more Fourier modes in the speckle pattern alias to the same point in ${\cal B}_1$, then prior knowledge is required to assign the correct amount of power to each mode. For the double slit setup, aliasing occurs if the fringe spatial frequency $u_{\rm fringe} = d_{\rm slit}P/z_0\lambda$ is near $\frac12$, 1, $\frac32$, etc. This would require more substantial changes to the current implementation.)

The same ``observed'' power spectrum can then be used to approximate $C_{ij}$:
\begin{equation}
C_{ij} \approx \int_{{\cal B}_1} P_{\rm obs}({\boldsymbol u}) 
\left[1
+ \sum_{\alpha=0}^{K-1} \xi_{i\alpha}
 T_\alpha^\ast({\boldsymbol u})
+ \sum_{\alpha=0}^{K-1} \xi_{j\alpha}
 T_\alpha({\boldsymbol u})
+ \sum_{\alpha=0}^{K-1}\sum_{\beta=0}^{K-1} \xi_{i\alpha}\xi_{j\beta}
 T_\alpha^\ast({\boldsymbol u}) T_\beta({\boldsymbol u})
\right]
 e^{2\pi i{\boldsymbol u}\cdot({\boldsymbol r}_i - {\boldsymbol r}_j)}\, d^2{\boldsymbol u},
\label{eq:Cij-a}
\end{equation}
where we have defined the functions
\begin{equation}
T_\alpha({\boldsymbol u}) \equiv \frac{\tilde B_\alpha({\boldsymbol u})}{\tilde{\bar R}({\boldsymbol u})} = \left\{
\begin{array}{cll}
1 & & \alpha=0 \\
-iu_x & & \alpha=1 \\
-iu_y & & \alpha=2 \\
-\frac12u_x^2 & & \alpha=3 \\
-u_xu_y & & \alpha=4 \\
-\frac12u_y^2 & & \alpha=5.
\end{array}
\right.
\end{equation}
Again, as long as we are working with oversampled speckle fringe data, we may consider only ${\boldsymbol u}\in{\cal B}_1$.

Finally, in practical laboratory setups, the local power spectrum in an $N_{\rm side}\times N_{\rm side}$ patch may be different from that in the larger power spectrum estimation region $N_{\rm PS}\times N_{\rm PS}$ because of geometrical distortions (we will show some examples of this in Section~\ref{sec:sims}). This results in a linear gradient of the pixel offsets across the patch. We thus make a ``gradient removed'' version of the pixel offset map $\xi_{i\alpha}$ by subtracting the least squares fit linear function of ${\bf r}_i - {\bf r}_i({\rm ctr})$ from the map, where ${\bf r}_i({\rm ctr})$ is the position of the center of the patch. Both the raw and gradient removed maps are examined in what follows.

The prior matrix ${\boldsymbol\Sigma}$ was taken to be diagonal with standard deviations $\Sigma_{i\alpha,i\alpha}^{1/2}=0.1$ for the astrometric offsets ($\alpha = 1$ or 2) and $10^{-4}$ otherwise (i.e., the flat and 2nd moment offsets are essentially turned off).

\subsection{Computational aspects}
\label{ss:tricks}

The calculation of $G_{i\alpha}$ in Eq.~(\ref{eq:GP}) is very computationally expensive since there are $NK$ quantities to evaluate, each one containing operations on $N\times N$ matrices (recall: typically $N=64^2=4096$). Therefore, it is essential to find tricks to speed up all of these computations. We also need an efficient means of solving the coupled non-linear system of $NK$ equations.

We first consider computation of ${\bf C}$. In Eq.~(\ref{eq:Cij-a}), we require the integrals of the form
\begin{equation}
I_{i\alpha,j\beta} \equiv
\int_{{\cal B}_1} P_{\rm obs}({\boldsymbol u}) T^\ast_\alpha({\boldsymbol u}) T_\beta({\boldsymbol u})\,e^{2\pi i{\boldsymbol u}\cdot({\boldsymbol r}_i-{\boldsymbol r}_j)}\,d^2{\boldsymbol u}.
\label{eq:Iform}
\end{equation}
In principle, there are $N^2K^2$ of these integrals (for each choice of pixels ${\boldsymbol r}_i$ and ${\boldsymbol r}_j$, and each choice of response basis function $\alpha$ and $\beta$), and if there are $N$ samples in the ${\boldsymbol u}$-plane, then by brute force one would need ${\cal O}(N^3K^2)$ operations. However, since the integral is needed on a regular grid ${\boldsymbol r}_i-{\boldsymbol r}_j\in{\mathbb Z}^2$ and can be computed by FFT, all of these integrals at all of the grid points can be obtained in ${\cal O}(NK^2\log N)$ operations.

The matrix inverse $\hat{\bf C}^{-1}$ is computed explicitly here by a conventional ${\cal O}(N^3)$ algorithm. Techniques in the CMB literature exist for iterative methods that avoid explicitly representing $\hat{\bf C}$ in memory and doing fast system solution $\hat{\bf C}^{-1}{\boldsymbol d}$ (these were used for mapmaking for the {\slshape WMAP} satellite, \citealt{1996ApJ...458L..53W}; introduced into CMB power spectrum estimation by \citealt{1999ApJ...510..551O}; and applied to CMB lensing by \citealt{2007PhRvD..76d3510S}). We have chosen instead to work with $N=64\times 64$ ``patches.'' At this size, an $N\times N$ matrix in double precision occupies 134 MB of memory, and with standard {\tt numpy} routines \citep{numpy} matrix multiplication and inversion are feasible.

The partial derivatives $\partial{\bf C}/\partial\xi_{i\alpha}$ represent a bigger challenge: since there are $NK$ $\xi_{i\alpha}$'s and $N^2$ entries in ${\bf C}$, the full set of partial derivatives has $N^3K$ entries; this is 550 GB in double precision even for $K=1$. Instead, we take advantage of the sparseness pattern of $\partial{\bf C}/\partial\xi_{i\alpha}$: since $\xi_{i\alpha}$ is a property of pixel $i$, only row $i$ and column $i$ of $\partial{\bf C}/\partial\xi_{i\alpha}$ are non-zero. We write
\begin{equation}
\frac{\partial{\bf C}}{\partial\xi_{i\alpha}} = {\bf D}^{(i\alpha)}+ {\bf D}^{(i\alpha)\,\rm T},
~~
D^{(i\alpha)}_{jk} = \delta_{ij} 
\int_{{\cal B}_1} P_{\rm obs}({\boldsymbol u}) 
\left[
 T_\alpha^\ast({\boldsymbol u})
+ \sum_{\beta=0}^{K-1} \xi_{i\alpha}\xi_{j\beta}
 T_\alpha^\ast({\boldsymbol u}) T_\beta({\boldsymbol u})
\right]
 e^{2\pi i{\boldsymbol u}\cdot({\boldsymbol r}_i - {\boldsymbol r}_k)}\, d^2{\boldsymbol u}.
\end{equation}
Because of the Kronecker delta, we may compute just the matrix $D^{(i\alpha)}_{ik}$, which has $N^2K$ components. These can be expressed in terms of the integrals of the form Eq.~(\ref{eq:Iform}), so can be obtained in ${\cal O}(NK^2\log N)$ operations. We then note that the operations with $\partial{\bf C}/\partial\xi_{i\alpha}$ needed in Eq.~(\ref{eq:GP}) can be computed in a ``fast'' mode. For example, we need to be able to compute
\begin{equation}
\frac12{\boldsymbol v}^{\rm T}\frac{\partial{\bf C}}{\partial\xi_{i\alpha}}{\boldsymbol v} = 
\sum_{k=0}^{N-1}D^{(i\alpha)}_{ik}v_iv_k
\end{equation}
for a vector ${\boldsymbol v}$; in the form written here, this is $N^2K$ operations for {\em all} of the $i\alpha$. We also need, for a symmetric matrix ${\bf M}$,
\begin{equation}
\frac12{\rm Tr}\left[
{\bf M}\frac{\partial{\bf C}}{\partial\xi_{i\alpha}}\right]
=
\sum_{k=0}^{N-1}D^{(i\alpha)}_{ik}M_{ik},
\end{equation}
which is also $N^2K$ operations for {\em all} of the $i\alpha$. So we may compute all of the $G_{i\alpha}$, and the dominant computation time is the ${\cal O}(N^3)$ time for the $N\times N$ matrix inversions and matrix multiplications (which can be done once and then applied to all $i\alpha$).

We also need a procedure to find the solution to $G_{i\alpha}=0$. One's first thought is to use the Newton-Raphson method; this would involve the derivative:
\begin{equation}
\frac{\partial G_{i\alpha}}{\partial\xi_{j\beta}}
= 
\sum_{\rm apertures} \left\{
\frac12 \sum_{\mu=0}^{N_{\rm R}-1}({\boldsymbol d}^{(\mu)}-\bar{\boldsymbol d})^{\rm T}\hat{\bf C}^{-1} \frac{\partial{\bf C}}{\partial\xi_{i\alpha}\partial\xi_{j\beta}}\hat{\bf C}^{-1}({\boldsymbol d}^{(\mu)}-\bar{\boldsymbol d})
- \frac12(N_{\rm R}-1) {\rm Tr} \left[ \hat{\bf C}^{-1}\frac{\partial{\bf C}}{\partial\xi_{j\beta}}\hat{\bf C}^{-1}\frac{\partial{\bf C}}{\partial\xi_{i\alpha}} \right]\right\}
-[{\boldsymbol\Sigma}^{-1}]_{i\alpha,j\beta}.
\end{equation}
One feature of the Newton-Raphson method is that it can converge even with only an approximate derivative. If $\partial{\bf C}/\partial\xi_{i\alpha}$ varies slowly, then we can drop the first term in braces in comparison to the second. We further drop blocks in the Jacobian that are off-diagonal in the basis functions, $\alpha=\beta$. We thus write the iterative scheme:
\begin{equation}
\Delta \xi_{i\alpha} = \varepsilon \sum_{j=0}^{N-1} [{\boldsymbol F}^{(\alpha\alpha)\,-1}]_{ij} G_{j\alpha} +
\left\{ \begin{array}{lll} \varepsilon'\Delta \xi_{i\alpha}({\rm previous})
 & & {\rm included} \\ 0 & & {\rm not~included}
\end{array}\right.
,
\label{eq:dxi-iter}
\end{equation}
where $\varepsilon$ and $\varepsilon'$ are tunable convergence parameters (and $\varepsilon'$ can be turned on or off), and
\begin{equation}
F^{(\alpha\beta)}_{ij} =  \sum_{\rm apertures}
\frac12(N_{\rm R}-1) {\rm Tr} \left[ \hat{\bf C}^{-1}\frac{\partial{\bf C}}{\partial\xi_{j\beta}}\hat{\bf C}^{-1}\frac{\partial{\bf C}}{\partial\xi_{i\alpha}} \right]
+ [{\boldsymbol\Sigma}^{-1}]_{i\alpha,j\beta}.
\label{eq:F1}
\end{equation}
There are $(NK)^2$ entries in this matrix, although only the $N^2K$ entries with $\alpha=\beta$ are actually used. Each one is a trace of a product of four $N\times N$ matrices, so would be an ${\cal O}(N^5K)$ computation if done by brute force. The traces can be computed more efficiently by:
\begin{equation}
\frac12{\rm Tr} \left[ \hat{\bf C}^{-1}\frac{\partial{\bf C}}{\partial\xi_{j\beta}}\hat{\bf C}^{-1}\frac{\partial{\bf C}}{\partial\xi_{i\alpha}} \right] = P^{(\alpha)}_{ij}P^{(\beta)}_{ji}
+ [\hat{\bf C}^{-1}]_{ij}\sum_{k=0}^{N-1}
  P^{(\alpha)}_{ik} D^{(j\beta)}_{jk},
  ~~{\rm where}~~
  P^{(\alpha)}_{ij} = \sum_{k=0}^{N-1} D^{(i\alpha)}_{ik} [\hat{\bf C}^{-1}]_{kj}.
\end{equation}
The ${\bf P}^{(\alpha)}$ matrices can be obtained by matrix multiplication in ${\cal O}(N^3K)$ operations, and the multiplication with the $D$'s is a further ${\cal O}(N^3K^2)$ operations (only $N^3K$ if the $\alpha=\beta$ blocks are needed). This process can be implemented using {\tt numpy} matrix functions, and allows highly efficient computation of ${\bf F}$.

Finally, we comment on the choice of parameters $\varepsilon$ and $\varepsilon'$. If $G_{i\alpha}$ were exactly a linear function of the $\zeta_{j\beta}$, and ${\bf F}$ were the exact matrix of partial derivatives, then we could reach convergence in a single step with $\varepsilon=1$ and $\varepsilon'=0$. A better choice is motivated by linear convergence theory (see, e.g., Chapter 3 of \citealt{Young71}). The true matrix of partial derivatives $\partial G_{i\alpha}/\partial\xi_{j\beta}$ may be written as ${\bf FV}$, where ${\bf V}$ is another matrix. If ${\bf V}$ were constant (so that we can write the error $\xi_{i\alpha}-\xi_{i\alpha}^{\rm(best~fit)}$ as a linear combination of the eigenvectors of ${\bf V}$) and the $\varepsilon'$ term is turned off, then in every iteration, the coefficient of an eigenvector corresponding to the eigenvalue $\lambda$ of ${\bf V}$ changes by a factor of $1-\varepsilon\lambda$. If we want stability up to some maximum eigenvalue $\lambda_{\rm max}>1$, we would set $\varepsilon=2/\lambda_{\rm max}$. If we include the $\varepsilon'$ term for even-numbered iterations, then every 2 iterations, the error is multiplied by a factor $f(\lambda) = 1-(2\varepsilon+\varepsilon'\varepsilon)\lambda + \varepsilon\lambda^2$. We chose $\varepsilon=2.6833/\lambda_{\rm max}$ and $\varepsilon'\varepsilon=1.8334/\lambda_{\rm max}$, so that $|f(\lambda)|<1$ for $0<\lambda<\lambda_{\rm max}$ and so that $f(\lambda)$ has a minimum of $-0.8$ (errors in the solution corresponding to modes with $f(\lambda)<0$ alternate in sign every 2 iterations, but this choice ensures that their amplitude decreases by at least 20\% every 2 iterations). After some experimentation, we chose $\lambda_{\rm max}=6$ and 20 iterations total as our default.

\section{Simulations}
\label{sec:sims}

In order to illustrate the technique, we run a suite of simulations. We generate two sets of simulations with an injected distortion. The first uses the Fraunhofer diffraction approach using a Fast Fourier Transform (FFT) of the slit-plane electric field. This generates a pattern of fringes whose power spectrum is truly stationary across the array. This simulation deviates from the setup because with a detector at a finite distance from the slit, there are geometric distortions of the fringe pattern (e.g., fringe spacings contain factors of $\sec^3\theta$ on the $x$-axis and $\sec\theta$ on the $y$-axis, where $\theta=0$ is normal incidence). The second uses the Fresnel diffraction integrals to produce a set of fringes with the correct geometric distortion and other non-far-field effects.

\begin{figure}
    \centering
    \includegraphics[width=6.5in]{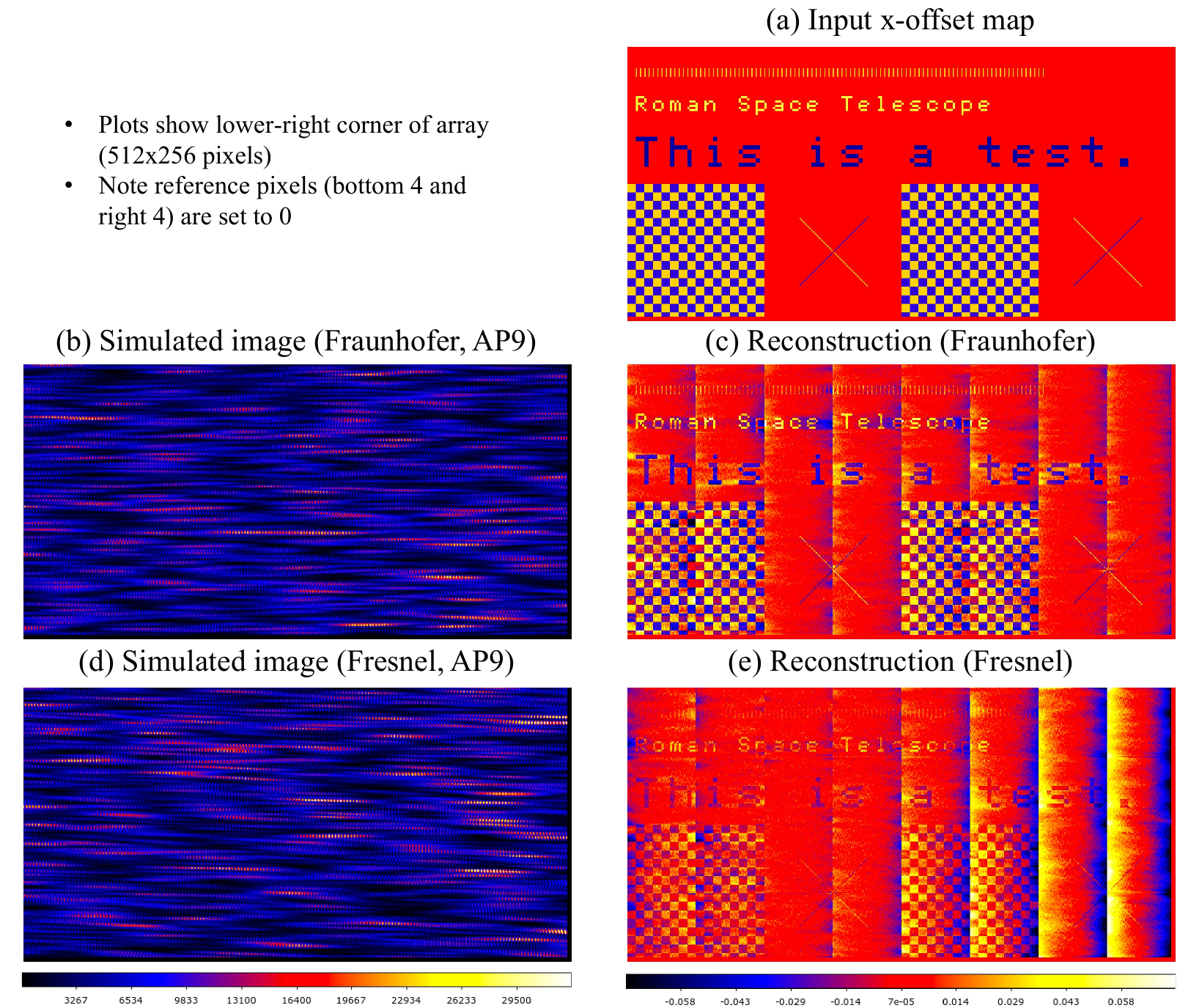}
    \caption{\label{fig:test}Some examples of the simulations from Section~\ref{sec:sims}. Each panel shows the lower-right $512\times 256$ region (i.e., {\tt[:256,-512:]} in {\sc Python} indexing) of the array. Panel (a) shows the input $x$-offset map, which contains several test patterns. Panel (b) shows a simulated image through the AP9 aperture using the Fraunhofer method (which produces a stationary pattern), including the indicated $x$-offsets. The reconstruction algorithm, applied in $64\times 64$ patches to 12 realizations each of AP4...AP9, produces the pattern in panel (c). Panels (d) and (e) show a simulation and reconstruction for the Fresnel simulation. In panel (e), the variation in the spatial frequency of the fringes results in the reconstruction systematically moving the pixels closer together (so positive $x$-offset in the left part of each $64\times 64$ patch and negative $x$-offset in the right part).}
\end{figure}

Each of our simulations has a test pattern for the $x$-offset map $\xi_{i,1}$, which consists of 512 repititions of a test pattern (Fig.~\ref{fig:test}a). We also set the 2nd moment in the simulation to $\xi_{i,3}=\xi_{i,1}^2$, since this is the proper Taylor expansion of $R_i(x+\xi_{i,1},y)$ in Eq.~(\ref{eq:RBasis}). (Another way to think about this is that when a pixel is offset, its non-reduced 2nd moment contains the centroid offset squared.)

\subsection{Fraunhofer method}
\label{ss:fraunhofer}

The Fraunhofer diffraction model treats the complex electric field at the detector as the Fourier transform of that in the slit plane:
\begin{equation}
E_{\rm det}({\boldsymbol r}) \propto \int E_{\rm slit}({\boldsymbol q}) e^{-2\pi i P{\boldsymbol r}\cdot{\boldsymbol q}/(\lambda z)}\,d^2{\boldsymbol q},
\label{eq:diff1}
\end{equation}
where ${\boldsymbol r}$ is a 2D vector measured in units of the pixel pitch, $z$ is the slit-detector spacing, and ${\boldsymbol q}$ is the 2D position vector in the slit plane. The slit is pixelized on an $N_{\rm slit}\times N_{\rm slit}$ grid, with a pixel spacing of $n_{\rm ov}\lambda z/(PN_{\rm slit})$, so that the FFT yields $E_{\rm det}$ on a grid with spacing $1/n_{\rm ov}$. The boundary effects in the FFT make the pattern periodic with period $N_{\rm slit}/n_{\rm ov}$ pixels. We thus set $N_{\rm slit} = n_{\rm ov}(N_{\rm side} + N_{\rm pad})$, where $N_{\rm pad}$ is some number of pixels added to prevent correlations from wrapping around the edges of the detector array. For our fiducial simulation, we set $n_{\rm ov}=8$ and $N_{\rm pad}=128$.

The map $E_{\rm slit}$ is set equal to zero except within the two slits. In the region within the slits, each pixel is set to a random 2D unit complex Gaussian (i.e., real and imaginary parts independent, each with mean 0 and variance 1), as one expects from the central limit theorem when one superposes many random reflection paths within the integrating sphere. The intensity at the detector plane is then $I({\boldsymbol r}) = |E_{\rm det}({\boldsymbol r})|^2$. We normalize the intensity to the mean value of the real ``half'' ramp dataset, which is 8292, 8335, 8550, 7923, 8006, and 8102 e/p (electrons per pixel) for apertures AP4 through AP9, respectively.

The intensity pattern is then convolved with a charge diffusion kernel (taken to be a Gaussian of width $\sigma_{\rm cd} = 0.33$ pix RMS \citep{2022PASP..134a4001G} and the pixel tophat. Optional pixel centroid offsets and second moments can be added at this stage, using Eq.~(\ref{eq:RBasis}), and using the fact that each basis function $\tilde B_\alpha({\boldsymbol u})$ can be obtained by multiplying by the appropriate polynomial in ${\boldsymbol u}$ before the FFT. Then we add Poisson noise; convolve with an inter-pixel capacitance kernel with $\alpha=0.0162$ (as measured from applying the tools of \citealt{2020PASP..132g4504F} to the acceptance test data); add read noise of $\sigma_{\rm read}=18$ e RMS (although this is small enough it has little effect); and divide by the gain ($g=1.697$ e/DN) to convert to simulated data numbers.

An example of a simulated image is shown in Figure~\ref{fig:test}b. The reconstruction is shown in Figure~\ref{fig:test}c. Note that each type of test feature (the checkerboard, cross, text, and barcode) appears in the reconstruction; but there is also some horizontal banding (since this fringe pattern gives very little constraint on the relative horizontal offsets of different rows) and some apodization (since the average is by construction zero, e.g., see the negative ``shadow'' underneath the positive ``Roman Space Telescope'' text).

\subsection{Fresnel method}
\label{ss:fresnel}

To make a more accurate simulation, we use the Fresnel integral,
\begin{equation}
E_{\rm det}({\boldsymbol r}) \propto \int E_{\rm slit}({\boldsymbol r}')
 \frac{z}{s^2} e^{2\pi is/\lambda}\,
d^2{\boldsymbol r}',
~~~
s = \sqrt{{P^2(\boldsymbol r}-{\boldsymbol r}')^2+z^2}
\label{eq:diff2}
\end{equation}
(see, e.g., \citet{1999poet.book.....B}, \S8.2, and with an inclination factor of $\cos\theta=z/s$; and recall that ${\boldsymbol r}$ and ${\boldsymbol r}'$ here are 2D vectors, so we add the 3rd dimension in the Pythagorean theorem separately). The complex exponential in Eq.~(\ref{eq:diff2}) uses the geometrical path length $s$ without making the linear approximation in Eq.~(\ref{eq:diff1}). Since $s$ is a function of ${\bf r}-{\bf r}'$, this is a convolution integral and can be implemented using FFT-based convolution methods. It is computationally intensive because of the large size of the FFTs necessary; we sample the grid at a $\delta = 5$ $\mu$m pitch. This discrete sampling of the Fresnel integral does lead to an artificial correlation in the diffraction at a separation of $z\lambda/(\delta P) = 3940$ pixels; but since the analysis of each $1024\times 1024$ patch is done independently (including determination of the power spectrum) this does not concern us. Memory usage peaked at $\approx 50$ GB and ``growing'' the FFTs by another factor of 2 on each axis would have required re-writing the simulation to run out of core.

The most noticeable result of the Fresnel integral is that the fringes are no longer uniformly spaced; instead there is a geometric distortion. The fringe spacing is smallest at the center of the array, and grows as one moves toward the edges. In the corners of the array, the fringes are sloped instead of horizontal. This is a result of geometric distortion when $\theta$ is not small and the small-angle approximation $\sin\theta\approx\theta$ breaks down.

Figure~\ref{fig:test}d shows a simulated image with the Fresnel method. There is a variation in the spatial frequencies of the fringes (narrower spacing on the left, wider spacing on the right), although it is not visible by eye. The reconstruction is shown in Figure~\ref{fig:test}e. The reconstruction attempts to move all the fringes to the mean spacing in the $512\times 512$ region, so on the right side it tries to move the pixels closer together. This produces the characteristic pattern where there is a positive reconstructed $x$-offset on the left side of each $64\times 64$ patch, and a negative reconstructed $x$-offset on the right side. This is in addition to the features in the test image, which are still present.

\section{Application to speckle data}
\label{sec:application}

We now consider the application of the algorithm to the speckle data, including how we handle some of the unique features of this setup. The application is to SCA 21536, one of the detector arrays from the {\slshape Roman} flight production lots that was not among the 18 selected to fly. 

The speckle projection system does not have a flat field, so we obtained the flat field $F_i$ from an average of 10 flats obtained during the acceptance testing. We used flat fields from the difference of frames 1 and 16 in the flats, which gives a mean signal level of $4\times 10^4$ e (using the gain from the acceptance test as determined by {\sc Solid-waffle}). The Poisson error per pixel in the flats is then expected to be $1/\sqrt{10\cdot4\times 10^4} = 0.16\%$. The non-linearity correction at this signal level estimated by {\sc Solid-waffle} would be 6.5\%, but this has not been applied. Since the acceptance test was done in a different Dewar with different illumination setup than the speckle projection system, the large-scale flat illumination pattern may be different. We have therefore computed a ``flat ratio'' map $q$ of the speckle data to the acceptance test flat. To remove the speckles, we average $q$ over all the input exposures and over $16\times 16$ pixel regions on the detector array. The resulting image was then degraded to $64\times 64$ pixel regions using a median filter, and then a further $3\times 3$ median filter was applied to this image. The mean operation must be performed first: medians are robust, but for speckle images the median is very different from the mean.\footnote{For single polarization speckle data, at a particular point, the intensity has a $\chi^2$ distribution with 2 degrees of freedom (real and imaginary parts of the electric field). For this distribution, the median is 0.693 times the mean. If we average $N_{\rm R}$ realizations of the pattern, there are $2N_{\rm R}$ degrees of freedom and in the limit of large $N_{\rm R}$ the ratio of median to mean becomes $1-1/(3N_{\rm R})$. The ratio may be further suppressed by the finite size of the pixel.}

There may be linear combinations of the $\xi_{i\alpha}$ that are poorly constrained by the data and default to zero because of the prior. Simulations can address this, but may not have all of the features of the data. Therefore we have also generated a modified set of images based on the real data but with artificial offsets injected. We generate an offset map $\xi_i^{\rm offset}$ to inject, and then take the image $I$ and interpolate it to the position $(x+\xi^{\rm offset}(x,y),y)$:
\begin{eqnarray}
I_{\rm inj}(x,y) &=& I(x,y) + \frac{\sin(2\pi u_{\rm fringe}\xi^{\rm offset}(x,y))}{\sin(2\pi u_{\rm fringe})} \frac{I(x+1,y)-I(x-1,y)}{2}
\nonumber \\
&& + \frac{1-\cos(2\pi u_{\rm fringe}\xi^{\rm offset}(x,y))}{1-\cos(2\pi u_{\rm fringe})}
\left[\frac{I(x+1,y)+I(x-1,y)}{2} - I(x,y)\right]
,
\label{eq:Ixy-mod}
\end{eqnarray}
where $u_{\rm fringe} = d_{\rm slit}P/(\lambda z_0)$ is the fringe spatial frequency in cycles per pixel. This is better than standard polynomial interpolation between neighboring points for a function that has only some Fourier modes present.\footnote{One can show that if $I(x,y)$ is any function of the form
$A(y) +B(y)\cos(2\pi u_{\rm fringe}x) + C(y) \sin(2\pi u_{\rm fringe} x)$, then this formula gives the correct value of $I(x+\xi^{\rm offset}(x,y),y)$. The interpolation formula gives an error if the fringes are at an integer multiple of the Nyquist frequency, $u_{\rm fringe} = \frac12, 1, \frac32,...$; this is because with fringes at these spatial frequencies it is not possible to uniquely solve for $A(y)$, $B(y)$, and $C(y)$. If $u_{\rm fringe}$ is small, Eq.~(\ref{eq:Ixy-mod}) reduces to standard 3-point quadratic interpolation.} These modified images were generated using an offset $\xi^{\rm offset}(x,y)$ independently drawn for each pixel using a uniform distribution between $-0.025$ and $+0.025$ pixels. The ``answer key'' $\xi^{\rm offset}(x,y)$ was also saved.

\begin{figure}
    \centering
    \includegraphics[width=6.45in]{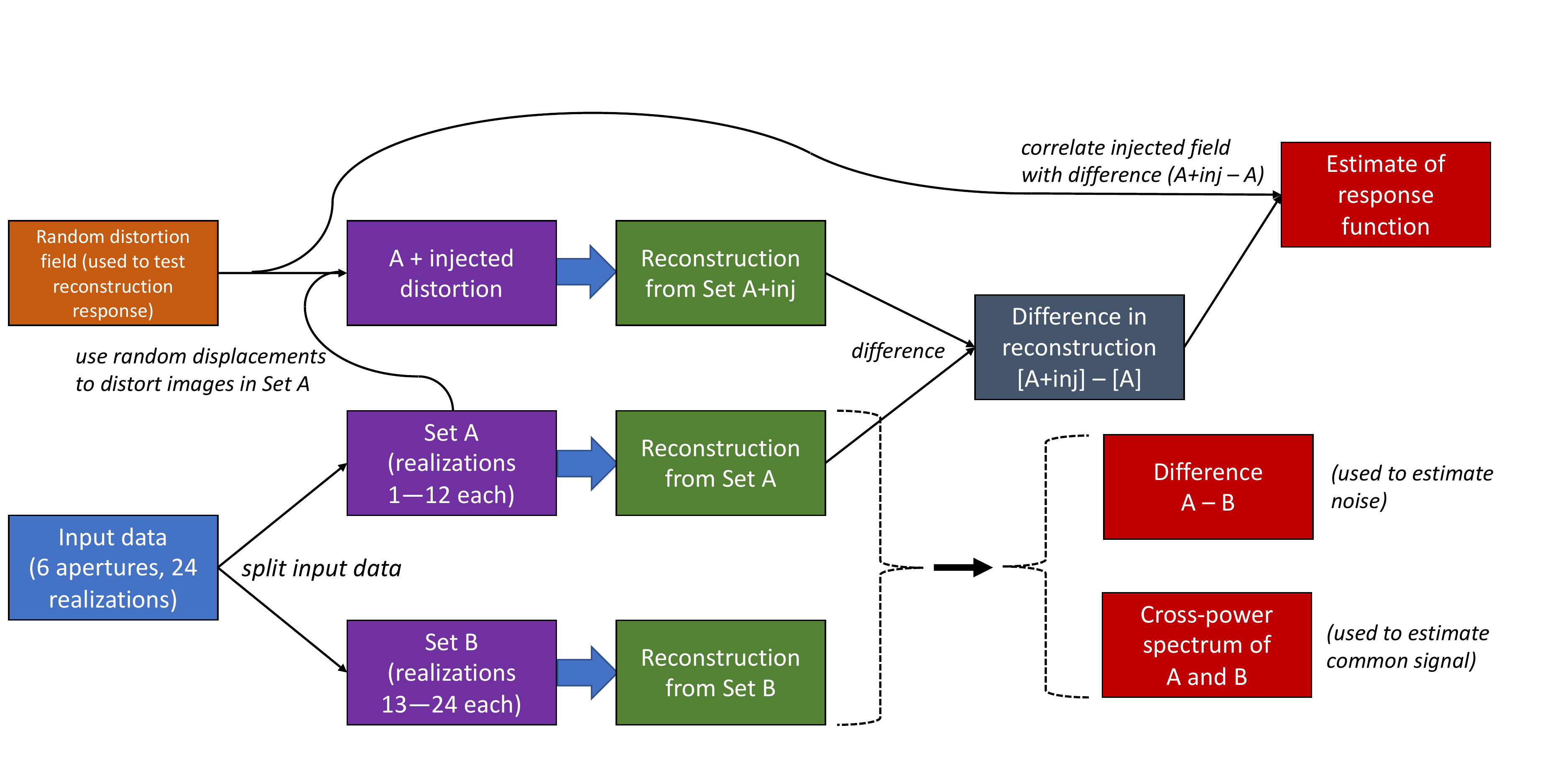}
    \caption{\label{fig:flowchart}The flow of injected signals and reconstructions used in Section \ref{sec:application}.}
\end{figure}

The flowchart of injected signals and reconstructions applied to the data is shown in Fig.~\ref{fig:flowchart}.

\subsection{Reconstruction runs}

Our fiducial reconstruction runs use the data acquired at wavelength 1.55 $\mu$m for apertures AP4 through AP8. We set tight priors (standard deviation $\Sigma_{i\alpha,i\alpha}^{1/2} = 10^{-4}$) to effectively ``turn off'' basis functions $\alpha = 0,3,4,5$, leaving only the astrometric offsets ($\alpha=1$ for $x$-offset and $\alpha=2$ for $y$-offset) to be measured. Weak priors ($\Sigma_{i\alpha,i\alpha}^{1/2} = 0.1$) are used for the pixel astrometric offsets.

The first step is to compress the data into a 2D image. The H4RG-10 can be non-destructively read; our exposures have 5 read frames each. We either perform our reconstruction on the ``full'' exposure (subtracting the 5th from the 1st read), or on a ``half'' exposure (subtracting the 3rd from the 1st read). The default choice is ``full.'' Due to the large signal level and the desire not to imprint additional correlations, we did not perform reference pixel subtraction.

For each output, we take the $4096^2$ detector array, and break it into 4096 patches that are each $64\times 64$ pixels (we denote $N_{\rm side}=64$, so we have a total of $N=N_{\rm side}^2 = 4096$ pixels per patch). This leads to smaller, more manageable matrix operations (recall many of the operations are order $N^3$, so reconstruction on $128\times 128$ patches would require $16\times$ more computing time). Pixels where the flat field $F_i$ is more than 50\% away from the median for that output channel are masked. We then ``enlarge'' the mask by masking a pixel if any of its 8 nearest neighbors (horizontal, vertical, or diagonal) are masked; this turned out to be necessary since pixels with such defects couple to their neighbors via inter-pixel capacitance \citep[e.g.][]{2004SPIE.5167..204M}, or (in the case of disconnected pixels) charge flowing into a neighbor.

Speckle reconstruction, like lensing reconstruction of the CMB, is inherently noisy. Therefore we split the speckle fringe images into an ``A'' set (realizations 1--12) and a ``B'' set (realizations 13--24). One can then determine how much of the structure in the reconstructed astrometric offsets is real and how much is noise by cross-correlating the A and B reconstructions. We generated three reconstructed outputs:
\begin{enumerate}
    \item ``A'': speckle realizations 1--12 for each aperture.
    \item ``B'': speckle realizations 13--24 for each aperture.
    \item ``A--inj'': a version of A but with an injected signal. Equation~(\ref{eq:Ixy-mod}) was applied to speckle realizations 1--12 for each aperture.
\end{enumerate}
We denote the output maps by $\hat\xi_\alpha(x,y)$, where $x$ and $y$ are integer pixel coordinates. Note that in the matrix operations, we used $\xi_{i\alpha}$, where $i$ is an index for a ``flattened'' array of pixels; for visualizing the output, it is easier to turn this back into a 2D image.

The cosmetic quality of SCA21536 is very good, and only 39082 of the active pixels (0.23\%) were masked.

\subsection{Response function}

We measure the response to injected signals as follows. For each patch, we compute the discrete 2D FFT:
\begin{equation}
\tilde \xi_{\alpha}(u_x,u_y) = \sum_{(x,y)\in\rm patch} \xi_\alpha(x,y) e^{-2\pi i(u_xx+u_yy)},
\end{equation}
where $u_x$ and $u_y$ are of the form integer/$N_{\rm side}$. We can then write the response to a particular Fourier mode by
\begin{equation}
R(u_x,u_y) = \frac{\sum_{\rm patches} \tilde \xi^{{\rm offset}\,\ast}(u_x,u_y) [\tilde{\hat \xi}_1(u_x,u_y|{\rm A-inj})-\tilde{\hat \xi}_1(u_x,u_y|{\rm A})]}{
\sum_{\rm patches} \tilde \xi^{{\rm offset}\,\ast}(u_x,u_y) \tilde \xi^{{\rm offset}}(u_x,u_y)}.
\label{eq:Ruxy}
\end{equation}
Here ``A--inj'' indicates that the reconstruction is performed from the A--inj input data. The response function is 1 if the input Fourier mode is faithfully represented in the output $\tilde{\hat \xi}_1(u_x,u_y|{\rm A-inj})$; it is 0 if the input Fourier mode has no effect on the output (i.e., if it has no effect on the data or is filtered out by the reconstruction); and in general it may be somewhere in between. Note that the reconstruction here is performed for $x$-astrometric offsets. Since a global astrometric offset cannot be measured by looking at a random field, we expect to have no response to the zero-mode: $R(0,0)=0$. This is in fact observed, although there is always some noise.

\begin{figure}
    \centering
    \includegraphics[width=6in]{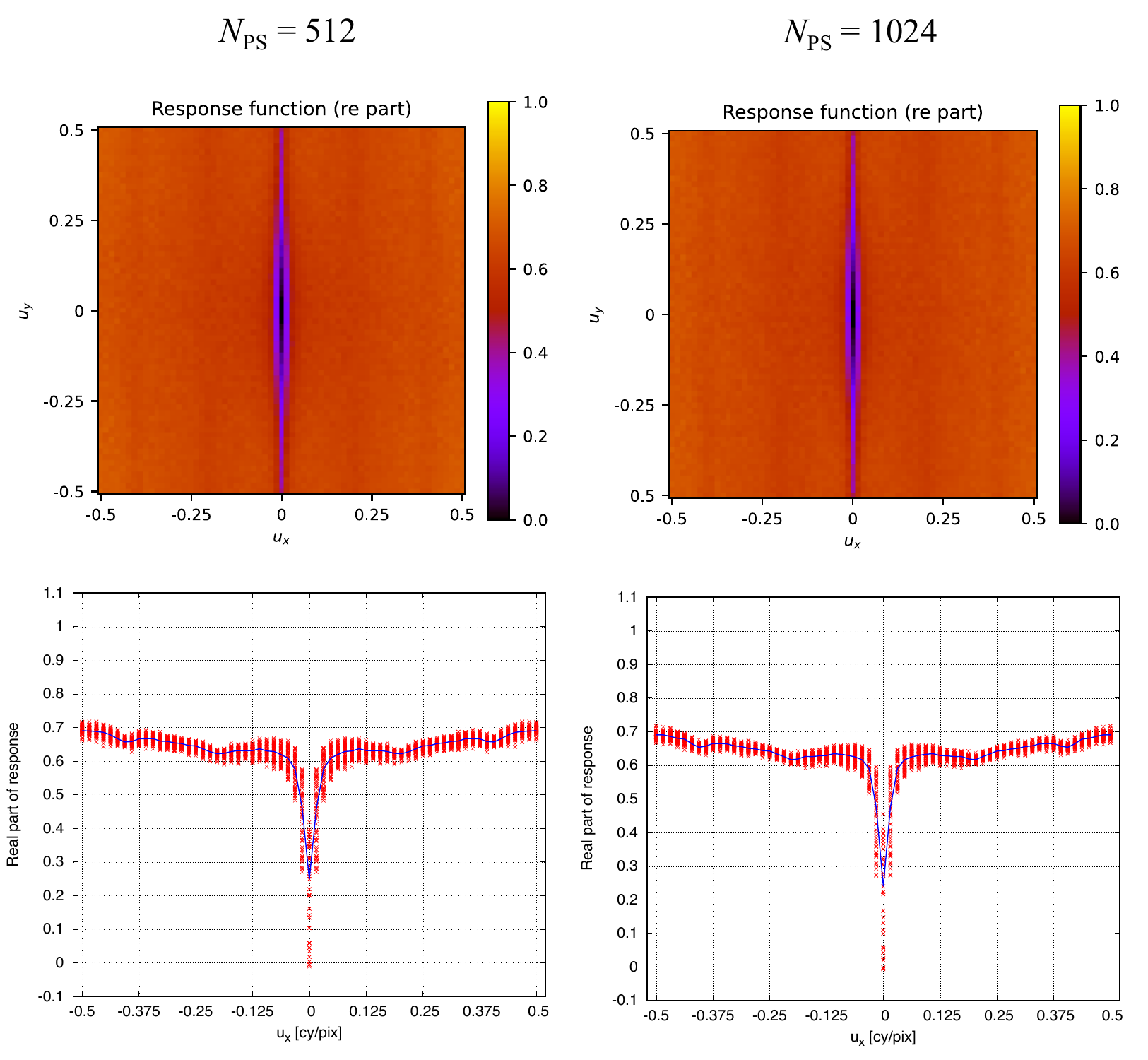}
    \caption{\label{fig:response}The response function $R(u_x,u_y)$ for 12 realizations of each of the AP4...AP8 apertures. The left part of each row shows results for the $N_{\rm PS}=512$ run and the right part for the $N_{\rm PS}=1024$ run. {\em Top}: A 2D map of the response function in the Fourier $(u_x,u_y)$ plane. Note the low response for modes with $u_x\approx 0$, i.e., Fourier wave vectors in the vertical direction. {\em Bottom}: A plot of the same data, but as a function of $u_x$ with each value of $u_y$ overplotted. The median for each $u_x$ is shown with the blue line.}
\end{figure}

The response function is shown in Figure~\ref{fig:response}. The median is 0.645 ($N_{\rm PS}=512$) or 0.643 ($N_{\rm PS}=1024$), indicating that pixel offsets injected into the data are recovered with a scaling factor of 0.643 (i.e., an offset of 0.01 pixels is reconstructed as 0.00643 pixels). The most visible feature in the response function map is that Fourier modes with $u_x\approx 0$ (i.e., wave vector in the vertical direction) are not well reconstructed. This makes sense: these modes correspond to complete rows in the image that are shifted left and right relative to each other. The speckle fringe pattern produces equally spaced fringes in the horizontal direction, enabling precise determination of the relative $x$-positions of pixels in the same row, but it de-correlates rapidly in the vertical direction so there is no information on the relative positions of pixels more than a few rows apart. The relative offsets between the different rows default to zero because of the prior, leading to near zero response function.

\subsection{Results for pixel offsets in SCA 21536}

\begin{figure}
    \centering
    \includegraphics[width=\textwidth]{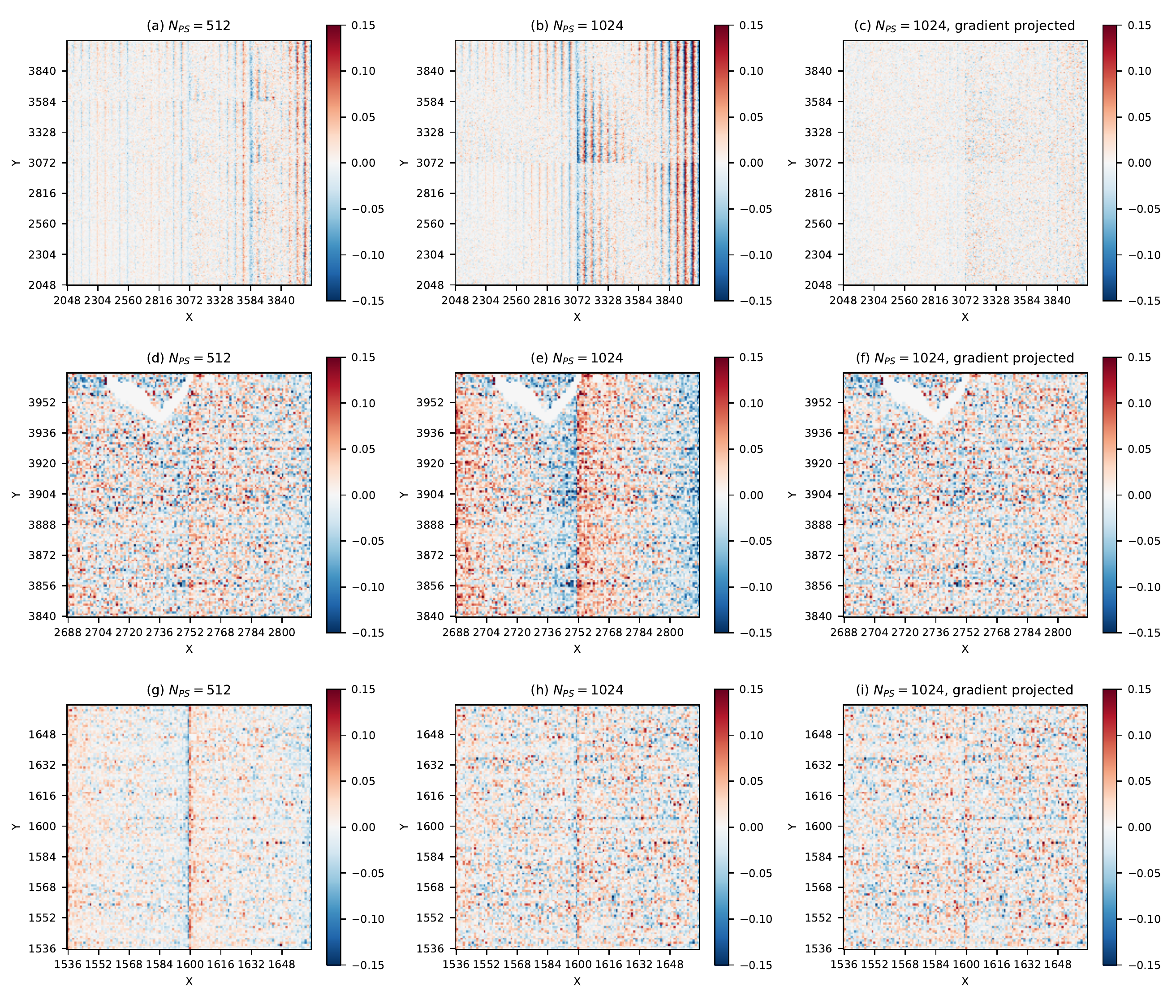}
    \caption{\label{fig:quadrant}Examples of the reconstructed pixel offsets. The map shows the pixel offsets in the $x$-direction (so red regions are pixels offset to the right and blue are pixels offset to the left). They are constructed by averaging the A and B reconstructions, and dividing by the median response (so (A+B)/(2$\times$0.64). All of the reconstructions are noise-dominated. Each row shows the reconstructions with two different region sizes for the power spectrum estimation ($N_{\rm PS}=512$ and 1024), and the last column shows the $N_{\rm PS}=1024$ region with the gradient projection. {\em Top row}: The upper-right quadrant of the SCA, smoothed $4\times 4$ for clarity. Note the gradient across each $64\times 64$ patch, which is more prominent in the case where the power spectrum is measured over the largest region ($N_{\rm PS}=1024$). {\em Middle row}: A zoom-in of a $128\times 128$ region, with no smoothing, containing a masked cosmetic defect (the V-shaped feature at top). Note that the noise in the map is correlated in horizontal bands since the speckle fringe pattern does not provide a good measurement of horizontal offsets of one row relative to another. {\em Bottom row}: Another $128\times 128$ region, without such a defect.}
\end{figure}

We now explore the results of the pixel offset characterization. Examples of the reconstructions are shown in Fig.~\ref{fig:quadrant}. For each panel in the figure, we show the average of the A and B reconstructions, divided by the median response function. This way the maps can be interpreted as maps of pixel offsets without multiplicative factors, with the caveats that (i) they are noise-dominated (i.e., most of the structure seen is noise rather than actual signal), and (ii) the modes with $u_x\approx 0$ have low response function and are missing.\footnote{One can think of this as very much like the missing spatial frequency problem in images reconstructed from radio interferometry (see \citealt{2017isra.book.....T}, in particular \S\S5.4 and 11.5.2).}

The top row of Fig.~\ref{fig:quadrant} shows the upper-right quadrant of the detector array, smoothed with a $4\times 4$ binning for clarity. The left image (panel a) shows the offsets in the $N_{\rm PS}=512$ reconstruction. One can see gradients across each $64\times 64$ patch (these correspond to colors that have a gradient from blue to white to red as one goes across the patch). These gradients result from the local fringe spacing being different from the $512\times 512$ region used to estimate the power spectrum $P_{\rm obs}({\boldsymbol u})$; the algorithm sees this different fringe spacing and tries to assign it to pixel offsets. One can see that there are 16 power spectrum estimation regions, and the gradient effect varies smoothly over each one. It is most pronounced on the right side of the figure, since that is where the fringe spacing has the strongest variation. These same effects are visible in the top-center panel (b), where we measure the power spectrum in $1024\times 1024$ regions; as one might imagine, the gradient effect is stronger in this case. In the right column (panel c), we have projected out the gradient (2 numbers: $x$ and $y$ slope) from each $64\times 64$ patch. This effectively removes the gradient pattern.

The middle row of Fig.~\ref{fig:quadrant} shows a zoomed-in $128\times 128$ pixel region at the native pixel resolution. The region chosen is from the top of the array ($3840\le y<3968$, $2688\le x<2816$), and was chosen to contain a large cosmetic defect. In panel (d), with the reconstruction at $N_{\rm PS}=512$, one can see the defect (the V-shaped feature, which was masked by our algorithm and thus set to all zeros). At this resolution, one can also see horizontal banding: the reconstruction algorithm is very good at measuring the relative positions of pixels within a row, but because of the pattern used, the position of each row relative to the next is very noisy. Panel (e) shows the same for $N_{\rm PS}=1024$, where the gradient effect is much more pronounced; this is effectively removed by the gradient projection (panel f), but the horizontal banding remains. The standard deviation of the map in panel (f) is $\sigma_x = 0.045$.

The final row of Fig.~\ref{fig:quadrant} shows another region ($1536\le x<1664$, $1536\le x<1664$), with no such defect. The standard deviation of the map in panel (i) is $\sigma_x =0.037$.

\subsection{Correlations of the A and B maps}
\label{ss:cf}

Since the pixel offset maps are dominated by noise, we use the cross-correlation of the ``A'' and ``B'' maps (obtained using different speckle realizations) in order to measure the power in the common signal. We show several measures of the A$\times$B correlation in Table~\ref{tab:summary}. All results are shown both with and without the gradient projected out, and for different regions of the Fourier (${\boldsymbol u}$) plane. The first group of results shows the raw covariance of the A and B maps:
\begin{equation}
{\rm Cov(A,B;uncorr.)} = \frac1{N_{\rm side}^2} \sum_{{\boldsymbol u}\,\rm included} \tilde\xi_1(u_x,u_y|{\rm A}) \tilde\xi_1^\ast(u_x,u_y|{\rm B}).
\end{equation}
The $1/N_{\rm side}^2$ comes from the usual normalization of the discrete Fourier transform; if one sums over all distinct ${\boldsymbol u}$ then one obtains $\langle \xi_1({\rm A})\xi_1({\rm B})\rangle$, and if one excludes the zero mode then one obtains the covariance ${\rm Cov}[ \xi_1({\rm A}),\xi_1({\rm B})]$.
The second group of results is corrected for the response function, 
\begin{equation}
{\rm Cov(A,B;corr.)} = \frac1{N_{\rm side}^2} \sum_{{\boldsymbol u}\,\rm included} \frac{\tilde\xi_1(u_x,u_y|{\rm A}) \tilde\xi_1^\ast(u_x,u_y|{\rm B})}{[R(u_x,u_y)]^2}.
\label{eq:cov-corr}
\end{equation}
The third group of results is the corrected covariance re-interpreted as an RMS, i.e., we take the square root of Eq.~(\ref{eq:cov-corr}). This interpretation is appropriate since if we have two maps A and B that contain a common signal $S$ and two independent noise fields $N_{\rm A}$ and $N_{\rm B}$, then the covariance of the A and B maps is equal to the variance of $S$. We also show the $1\sigma$ statistical uncertainty on this ``RMS correlated signal,'' computed from 1000 bootstrap re-samplings of the 4096 patches.\footnote{For stability, the bootstrap was computed on the corrected covariances, and then propagated to the RMS correlated signal using the usual derivative rule, $\sigma_{\sqrt x}=\sigma_x/(2\sqrt x)$.}

\begin{figure}
    \centering
    \includegraphics[width=\textwidth]{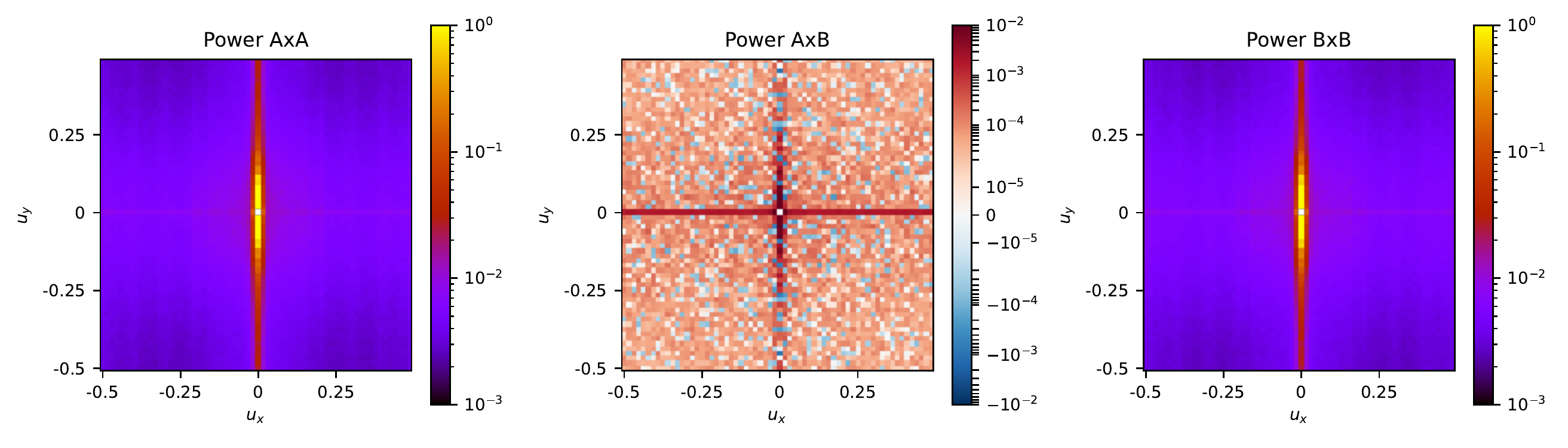}
    \caption{\label{fig:cross-power}The power spectra of the reconstructed pixel offsets for the case of $N_{\rm PS}=512$ with gradient removal. The power spectra are normalized so that $\int_{{\cal B}_1} P({\bf u})\, d^2{\bf u}$ is the variance (or covariance in the case of the A$\times$B cross power spectrum).}
\end{figure}

Of most direct interest to us is the RMS correlated signal. We focus on the gradient-removed case. There is a common signal of 0.043 pix RMS ($N_{\rm PS}=512$) or 0.110 pix RMS ($N_{\rm PS}=1024$), but almost all of this is at $u_x=0$, i.e., it corresponds to horizontal offsets of one row relative to another. These horizontal offsets of entire rows are difficult to constrain with our speckle pattern. If we exclude those modes -- i.e., take only the $u_x\neq 0$ modes -- the RMS correlated signal drops down to 0.012 pix RMS (full exposure). This result is also almost independent of $N_{\rm PS}$ (it varies by only 4\% when $N_{\rm PS}$ changes by a factor of 2). Of this signal, 0.009 pix RMS is coming from modes with $u_x\neq 0$ but $u_y=0$ (corresponding to columns that are not equally spaced), and 0.008 pix RMS is coming from modes with $u_x\neq 0$ and $u_y\neq 0$ (i.e., modes that are not overall offsets of rows or of columns). The contribution of different parts of the Fourier plane to the correlation of the maps can also be displayed as a 2D cross-correlation power spectrum, as shown in Fig.~\ref{fig:cross-power}.

The results for the half exposure are similar to the full exposure: the RMS correlated signal at $u_x\neq 0$ (for example) is 0.0107 pix RMS for the half exposure versus 0.0116 pix RMS for the full exposure. The difference is statistically significant when we average down over the full array. Recall that this test was done to assess the impact of non-linear effects (e.g., classical non-linearity, brighter-fatter effect) on our pixel offset measurements. The difference of only 8\%\ indicates that non-linearity is not a major driver of the correlated signal that we find.

We have also tried removing a quadratic function (also including $x^2$, $xy$, and $y^2$ terms in the fit) instead of a gradient. The RMS correlated signal in the $u_x\neq 0$ pixel offsets decreases from 0.0116 to 0.0106 ($N_{\rm PS}=512$, full exposure) or 0.0107 to 0.0097 ($N_{\rm PS}=512$, half exposure).

\def\z{\textcolor{white}0}
\begin{table}[]
\footnotesize
\centering
    \begin{tabular}{c|c|ccc|ccc}
    \multicolumn8c{Summary of A$\times$B correlations in pixel offset maps}  \\ \hline
    Modes included & Covariance & \multicolumn3c{Covariance} & \multicolumn3c{RMS correlated signal} \\
    & (uncorrected) & \multicolumn3c{(corrected)} & & & \\
        &  $N_{\rm PS}=512$ & $N_{\rm PS}=512$ & $N_{\rm PS}=512$ & $N_{\rm PS}=1024$ & $N_{\rm PS}=512$ & $N_{\rm PS}=512$ & $N_{\rm PS}=1024$ \\ 
        &  full & half & full & full & half & full & full \\ \hline
    \multicolumn8c{Gradient not removed} \\ \hline
    $(u_x,u_y)\neq (0,0)$ & $1.32\times 10^{-4}$ & $1.27\times 10^{-3}$ & $2.39\times 10^{-3}$ & $1.01\times 10^{-2}$ & 0.036\z\z$\pm$0.033\z\z\! & 0.049\z\z$\pm$0.020\z\z\! & 0.101\z$\pm$0.021\z \\
    $u_y\neq 0$ & $2.47\times 10^{-5}$ & $6.00\times 10^{-4}$ & $1.63\times 10^{-3}$ & $8.22\times 10^{-3}$ & 0.024\z\z$\pm$0.048\z\z\! & 0.040\z\z$\pm$0.024\z\z\! & 0.091\z$\pm$0.023\z \\
    $u_x\neq 0$ & $1.32\times 10^{-4}$ & $7.26\times 10^{-4}$ & $8.25\times 10^{-4}$ & $1.98\times 10^{-3}$ & 0.0270\z$\pm$0.0003\z\! & 0.0287\z$\pm$0.0004\z\! & 0.0445$\pm$0.0007 \\
    $u_x\neq 0$ and $u_y=0$ & $1.08\times 10^{-4}$ & $6.70\times 10^{-4}$ & $7.60\times 10^{-4}$ & $1.91\times 10^{-3}$ & 0.0259\z$\pm$0.0004\z\! & 0.0276\z$\pm$0.0004\z\! & 0.0437$\pm$0.0007 \\
    $u_x\neq 0$ and $u_y\neq 0$ & $2.43\times 10^{-5}$ & $5.67\times 10^{-5}$ & $6.50\times 10^{-5}$ & $7.22\times 10^{-5}$ & 0.0075\z$\pm$0.0001\z\! & 0.0081\z$\pm$0.0002\z\! & 0.0085$\pm$0.0002 \\ \hline
    \multicolumn8c{Gradient removed} \\ \hline
    $(u_x,u_y)\neq (0,0)$ & $3.85\times 10^{-5}$ & $1.31\times 10^{-3}$ & $1.88\times 10^{-3}$ & $1.22\times 10^{-2}$ & 0.036\z\z$\pm$0.023\z\z\! & 0.043\z\z$\pm$0.015\z\z\! & 0.110\z$\pm$0.014\z \\
    $u_y\neq 0$ & $2.41\times 10^{-5}$ & $1.25\times 10^{-3}$ & $1.81\times 10^{-3}$ & $1.21\times 10^{-2}$ & 0.035\z\z$\pm$0.023\z\z\! & 0.043\z$\pm$0.015\z\! & 0.110\z$\pm$0.014\z \\
    $u_x\neq 0$ & $3.82\times 10^{-5}$ & $1.15\times 10^{-4}$ & $1.35\times 10^{-4}$ & $1.46\times 10^{-4}$ & 0.0107\z$\pm$0.0001\z\! & 0.0116\z$\pm$0.0001\z\! & 0.0121$\pm$0.0001 \\
    $u_x\neq 0$ and $u_y=0$ & $1.43\times 10^{-5}$ & $6.23\times 10^{-5}$ & $7.24\times 10^{-5}$ & $8.08\times 10^{-5}$ & 0.0079\z$\pm$0.0001\z\! & 0.0085\z$\pm$0.0001\z\! & 0.0090$\pm$0.0001 \\
    $u_x\neq 0$ and $u_y\neq 0$ & $2.38\times 10^{-5}$ & $5.28\times 10^{-5}$ & $6.25\times 10^{-5}$ & $6.55\times 10^{-5}$ & 0.0073\z$\pm$0.0001\z\! & 0.0079\z$\pm$0.0002\z\! & 0.0081$\pm$0.0002 \\
    \hline
    \multicolumn8c{Quadratic function removed} \\ \hline
    $(u_x,u_y)\neq (0,0)$ & $3.65\times 10^{-5}$ & $2.03\times 10^{-3}$ & $2.61\times 10^{-3}$ & $8.79\times 10^{-3}$ & 0.045\z\z$\pm$0.013\z\z\! & 0.051\z\z$\pm$0.008\z\z\! & 0.094\z$\pm$0.008\z \\
    $u_y\neq 0$ & $2.40\times 10^{-5}$ & $1.98\times 10^{-3}$ & $2.56\times 10^{-3}$ & $8.73\times 10^{-3}$ & 0.045\z\z$\pm$0.013\z\z\! & 0.051\z\z$\pm$0.008\z\z\! & 0.093\z$\pm$0.008\z \\
    $u_x\neq 0$ & $3.61\times 10^{-5}$ & $9.47\times 10^{-5}$ & $1.12\times 10^{-4}$ & $1.25\times 10^{-4}$ & 0.0097\z$\pm$0.0001\z\! & 0.0106\z$\pm$0.0001\z\! & 0.0112$\pm$0.0002 \\
    $u_x\neq 0$ and $u_y=0$ & $1.24\times 10^{-5}$ & $4.27\times 10^{-5}$ & $5.04\times 10^{-5}$ & $6.03\times 10^{-5}$ & 0.00654$\pm$0.00004\! & 0.00710$\pm$0.00005\! & 0.0078$\pm$0.0001 \\
    $u_x\neq 0$ and $u_y\neq 0$ & $2.37\times 10^{-5}$ & $5.19\times 10^{-5}$ & $6.43\times 10^{-5}$ & $6.16\times 10^{-5}$ & 0.0072\z$\pm$0.0001\z\! & 0.0079\z$\pm$0.0002\z\! & 0.0080$\pm$0.0002 \\
    \end{tabular}
    \caption{\label{tab:summary}A table of the summary statistics of the pixel offsets in the $x$-direction, as measured from the cross-correlation of the ``A'' and ``B'' reconstructions. There are three parts to the table: without removing the overall gradient from the reconstructions; removing the gradient (best fit 1st order polynomial); and removing a quadratic function (best fit 2nd order polynomial). Values are shown both for the $N_{\rm PS}=512$ and $N_{\rm PS}=1024$ power spectrum binnings, and for $N_{\rm PS}=512$ we show both the ``half'' and ``full'' exposure reconstructions. The first set of values is the covariance between the A and B reconstructions; the second set is corrected for the response function. The third set is the square root of the corrected covariance, i.e., the common mode signal measured as an RMS $x$-offset. The first row contains all Fourier modes (except for the ${\boldsymbol u}=0$ mode, which is simply a constant offset and does not affect relative positions of pixels). The subsequent rows contain only some of the Fourier modes. The first two rows (which include $u_x=0$ modes) have large uncertainties. Note that the geometric distortion pattern is in modes with $u_y=0$ and is common mode since it is present in both the A and B subsets of the data.}
\end{table}

\begin{figure}
    \centering
    \includegraphics[width=6.5in]{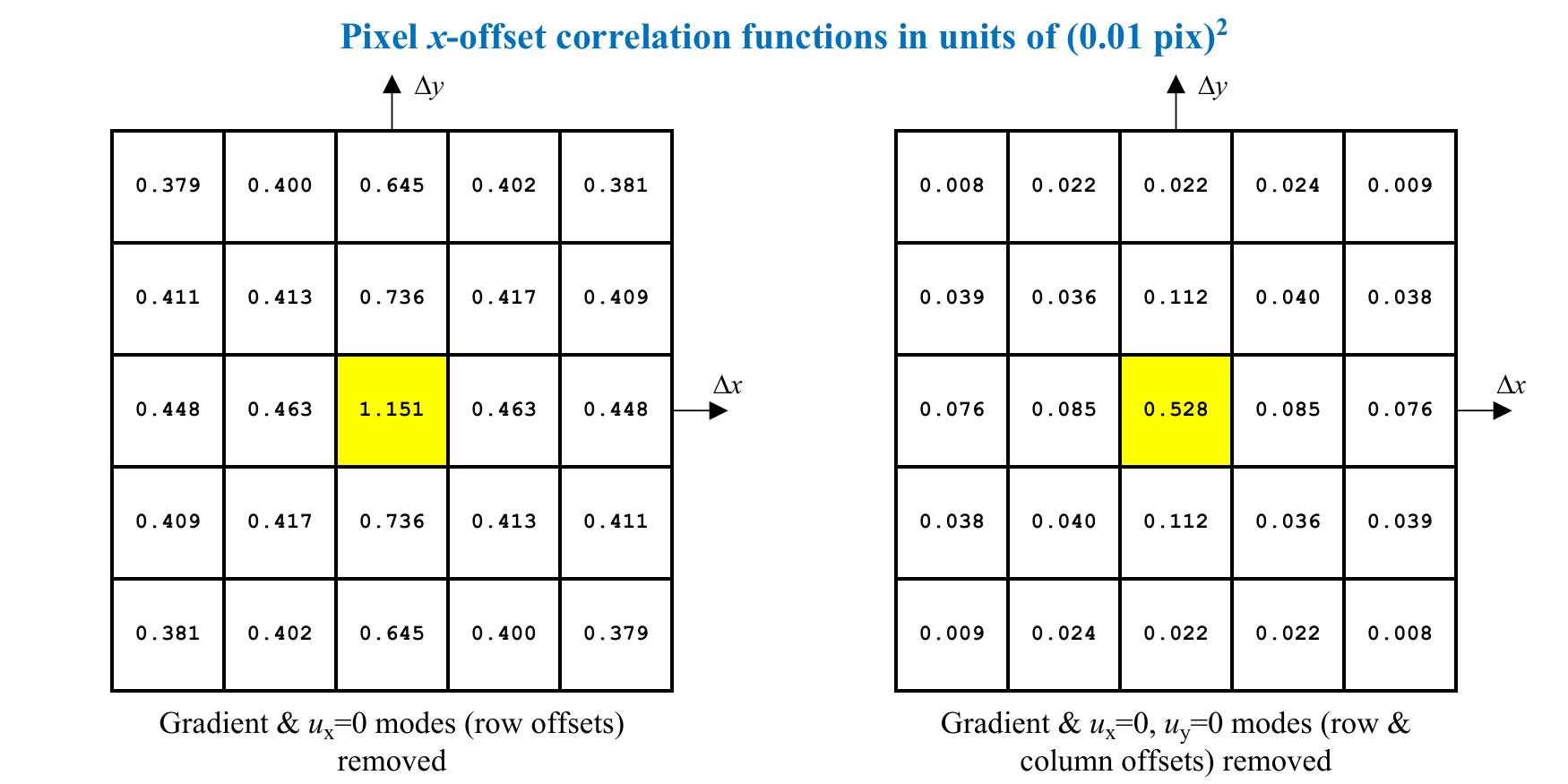}
    \caption{\label{fig:ccf}The correlation function of $x$-pixel offsets, Eq.~(\ref{eq:Cxx}), measured in units of $10^{-4}{\rm~pix}^2 = (0.01{\rm~pix})^2$. We show the $5\times 5$ region around zero lag ($\Delta x,\Delta y = -2...2$). The gradient-removed maps are used. The left panel shows the correlation function excluding the overall row offsets, and the right panel shows the correlation function excluding the overall column offsets as well. The $(0,0)$ pixel (highlighted in yellow) is equivalent to the RMS correlated signal in Table~\ref{tab:summary}, e.g., for the left panel $\sqrt{1.151}\times 0.01 = 0.0107$ pix RMS and for the right panel $\sqrt{0.528}\times 0.01 = 0.0073$ pix RMS.}
\end{figure}

Finally, we can investigate our results by measuring the correlation function of the $x$-pixel offsets in the A and B maps,
\begin{equation}
C_{xx}(\Delta x,\Delta y)
\equiv \langle \xi_1(x,y) \xi_1(x+\Delta x,y+\Delta y) \rangle
 = \frac1{N_{\rm side}^2} \sum_{{\boldsymbol u}\,\rm included} \frac{\tilde\xi_1(u_x,u_y|{\rm A}) \tilde\xi_1^\ast(u_x,u_y|{\rm B})}{[R(u_x,u_y)]^2} \cos [2\pi (u_x\Delta x + u_y\Delta y)]
 \label{eq:Cxx}
\end{equation}
(here the first equality is the definition of the correlation function, and the second is the method of computation used). By construction, $C_{xx}(0,0)$ is the covariance of the A and B maps corrected for the estimator response (see Eq.~\ref{eq:cov-corr}). The correlation functions are shown in Fig.~\ref{fig:ccf}, for both the $u_x\neq 0$ case (overall offset of each row removed) and the $u_x=0$ and $u_y=0$ case (overall offset of each column removed). We see that after the overall row and column offsets are taken out (right panel), the remaining centroid offsets appears to be dominated by individual pixel effects rather than groupings of a few pixels: the correlation coefficient between neighboring pixels is $C_{xx}(1,0)/C_{xx}(0,0) = 0.16$ (horizontal neighbors) or $C_{xx}(0,1)/C_{xx}(0,0) = 0.21$ (vertical neighbors).

\subsection{Summary}

We conclude that -- although the speckle position measurement is dominated by noise -- we can look for correlation between the different reconstructions to determine how much power in the positional offset maps is common between the A and B reconstructions. We focus our attention on relative offsets within each $64\times 64$ patch, and remove the overall horizontal offset of each row (i.e., we consider only $u_x\neq 0$ modes in Fourier space) since the fringe measurement has poor sensitivity to these offsets. We find a remaining residual correlated signal of 0.0107 pixels RMS if a first-order polynomial is removed from each patch ($N_{\rm PS}=512$, half exposure). This reduces to 0.0097 pix RMS if we fit a second-order polynomial. The results have only small changes (up to 15\%) if we use the full exposure or use larger regions to estimate the power spectrum of the speckle pattern ($N_{\rm PS}=1024$), indicating that they are robust against detector non-linearities and the geometric varying fringe spacing effect.

\section{Implications for PSF fitting}
\label{sec:psf}

The pixel-level variations in centroid position are a particular concern for fitting of the point-spread function (PSF), since the PSF is fit from stars with a high signal-to-noise ratio per object and hence the fit is dominated by signal from a small number of pixels. We have thus tested the impact of pixel centroid offsets by injecting them into the simulated images of stars and testing PSF recovery.

The PSF fitting simulation used is similar in spirit to those of \citet{2008PASP..120.1307M}, but with a few differences. Stars in a simulated exposure are generated with a magnitude distribution from the Trilegal v1.6 simulation at the South Galactic Pole \citep{2012ASSP...26..165G} and placed at random positions in the focal plane (including random sub-pixel position). The magnitudes of the stars are converted to electrons per exposure using the Exposure Time Calculator, v19 \citep{2012arXiv1204.5151H}; only stars with a total count between $10^4$ and $2.06\times 10^5$ electrons (so signal-to-noise ratio $>100$ but at less than half full well in the brightest pixel) are accepted. The ``base’’ model for the PSF includes FFT-based diffraction from the {\slshape Roman} pupil shape and the field-dependent Zernike coefficients from the Project website\footnote{URL: \tt http://roman.gsfc.nasa.gov/}. In order to only study the pixel centroid offsets in isolation, chromatic and polarization wavefront effects, finite detector thickness, inter-pixel capacitance, and the brighter-fatter effect were turned off. Charge diffusion of 3 $\mu$m rms per axis is included. The stars are pixelated at the 0.11 arcsec pixel scale of {\slshape Roman} (hence are undersampled) and Poisson noise and read noise ($\sigma_{\rm read}=7$ e RMS, effective over the length of the exposure after ramp fitting) are added. A $9\times 9$ postage stamp is cut out around each star and saved. We simulate 20 exposures, each with 160 stars (this is 1/5 of the 800 stars that are expected). This simulation is still many steps removed from a ``science-ready’’ implementation, but it is a useful testing ground to understand the importance of various physical effects.

We run the simulation both with and without pixel centroid offsets. For the simulation ``with’’ pixel centroid offsets, each postage stamp has a set of $x$-offsets generated according to the correlation function of Eq.~(\ref{eq:Cxx}). The PSF recovery algorithm does not know what pixel centroid offsets were injected.

Like \citet{2008PASP..120.1307M}, we fit a set of parameters in each exposure. We have used several different sets of parameters, but for the test presented here we used 10 parameters describing the state of the system: 5 global Zernike offsets (focus, astigmatism, and coma); 2 focus gradient offsets (essentially X and Y tilts of the focal plane); and 3 line-of-sight jitter covariance components. Unlike \citet{2008PASP..120.1307M}, we do not fit moments of the observed stars, but rather the postage stamps themselves. We then fit the target function:
\begin{equation}
\chi^2 = 2 \sum_{i} \left[ M_i - I_i - (I_i+\sigma_{\rm read}^2)\ln\frac{M_i +\sigma_{\rm read}^2}{I_i+\sigma_{\rm read}^2} \right],
\end{equation}
where the sum is over pixels in all of the postage stamps; $I_i$ is the observed intensity in pixel $I$; and $M_i$ is the model in pixel $i$. This $\chi^2$ function is that for a Poisson distribution offset by $\sigma_{\rm read}^2$ so that it has the correct variance. (Pixels with $I_i\le -\sigma_{\rm read}^2$, which can occur due to noise fluctuations, are set to $I_i = -\sigma_{\rm read}^2 + 10^{-24}$ to avoid numerical error in the logarithm.) We run a loop that alternates between fitting the fluxes and centroid coordinates of the stars (which can be done independently for each star for a fixed set of optical parameters) and updating the optical parameters.

\begin{figure}
\centering{
\includegraphics[width=3.2in]{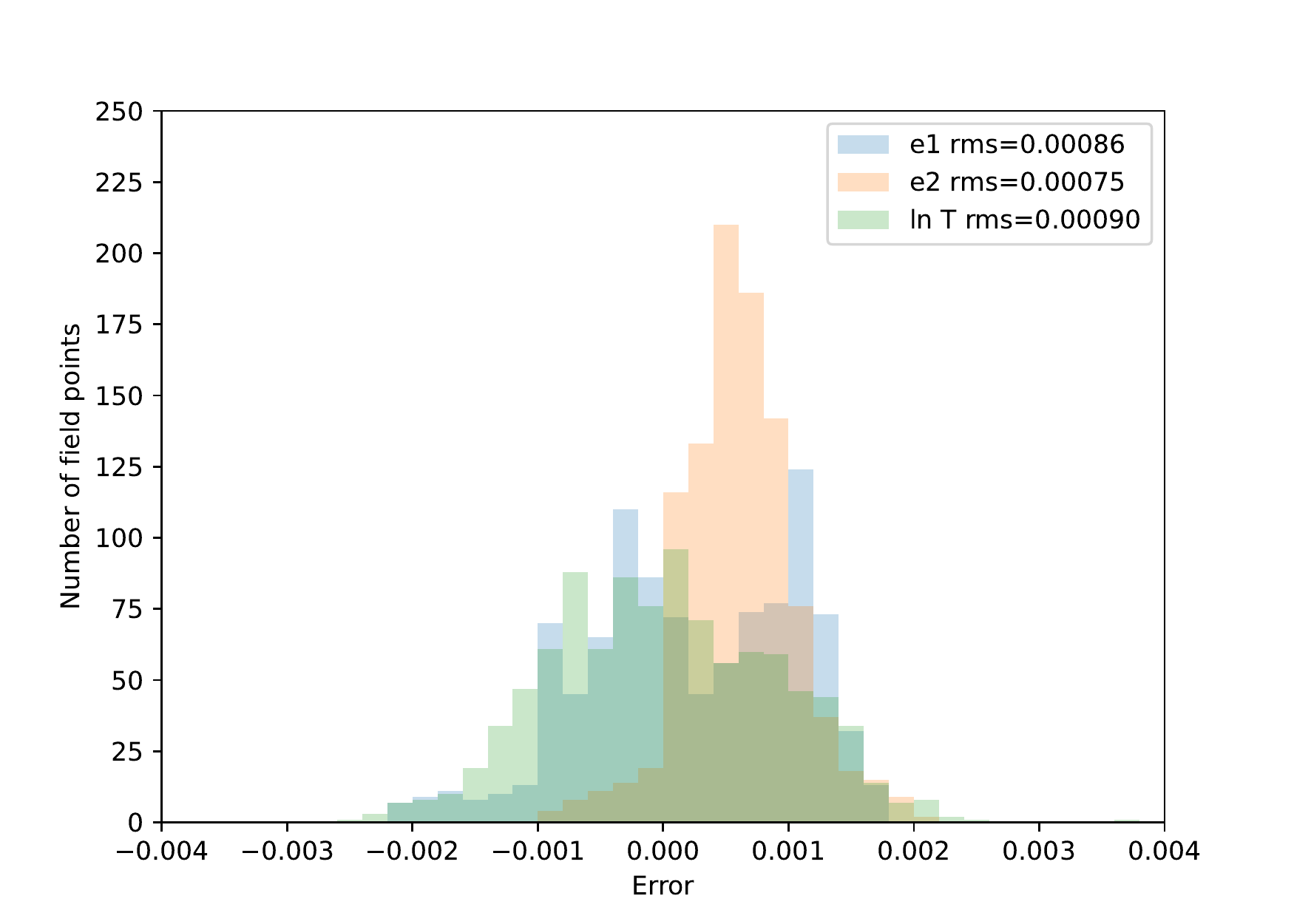}
\includegraphics[width=3.2in]{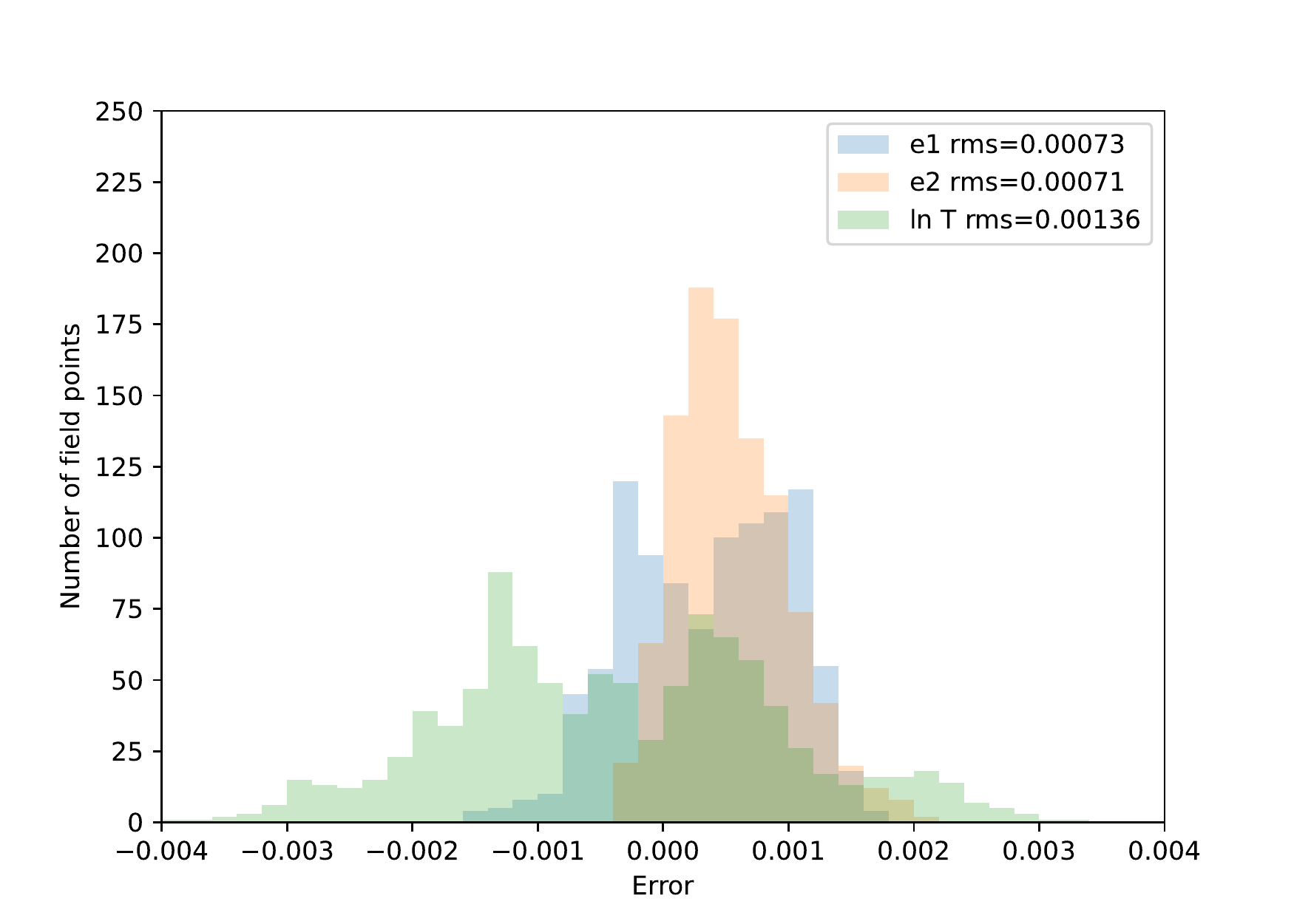}
}
\caption{\label{fig:psf-hist}The histogram of errors in simulated recovered PSF second moments $\Delta e_1$, $\Delta e_2$, and $\Delta \ln T$, for the case without the pixel centroid offsets (left), and with the pixel centroid offsets (right). Note that these simulations use only 1/5 of the available stars at South Galactic Pole density, and that for the case with pixel centroid offsets no corrections were applied (i.e., the model-fitting pipeline did not know that the centroid offsets were included).}
\end{figure}

For each of our 20 simulations, we selected 50 random field points and computed the model PSF, both from the ``truth’’ input model and from the best-fit from the $\chi^2$ procedure. We compute the $2\times 2$ matrix of adaptive second moments ${\bf M}$ \citep{2002AJ....123..583B}, and report the errors
$\Delta \ln T$, $\Delta e_1$, and $\Delta e_2$ in the log-trace and 2 ellipticity components:
\begin{equation}
\ln T = \ln(M_{xx}+M_{yy}), ~~~ e_1 = \frac{M_{xx}-M_{yy}}{M_{xx}+M_{yy}}, ~~~{\rm and}~~~ e_2 = \frac{2M_{xy}}{M_{xx}+M_{yy}}.
\end{equation}
The histograms of these errors are shown in Figure~\ref{fig:psf-hist}. Results are shown for the $J$ band, since it is the most demanding of the {\slshape Roman} shape measurement bands.

Figure~\ref{fig:psf-hist} also shows the RMS error in the PSF size and ellipticity, with and without the inclusion of the $x$-centroid offsets. We see that the inclusion of the $x$-centroid offsets had no measurable impact on the ellipticity errors (the RMS errors actually went down, but based on 20 simulations this is not statistically significant). On the other hand, the PSF size error increased by a factor of 1.7.

The {\slshape Roman} Science Requirements Document\footnote{Available on the Project website: \tt https://asd.gsfc.nasa.gov/romancaa/docs2/RST-SYS-REQ-0020C\_SRD.docx} specifies that the PSF ellipticity shall be known to $5.7\times 10^{-4}$ RMS per component, and the PSF size $\ln T$ shall be known to $7.2\times 10^{-4}$ RMS, after combining the 5 dither positions used in the Reference Survey. (See \citealt{2021MNRAS.501.2044T} for more details on the derivation of this requirement.) Given that we have used only 1/5 of the stars, and that Fig.~\ref{fig:psf-hist} is based on independent fits to individual exposures, the RMS error of 0.00136 (i.e., $>7.2\times 10^{-4}$) on $\ln T$ does not in and of itself imply that a correction for individual pixel offsets is necessary. However, it is clear that the pixel offsets are not a negligible part of the overall PSF error budget. Furthermore, one might expect that the inclusion of the other pixel-to-pixel variations that were not measured in this study, such as the $y$-centroid offsets or higher moments, would further increase the spread in $\ln T$. We intend to revisit this question after the next set of characterization tests using deterministic patterns.

\section{Discussion}
\label{sec:discussion}

We have carried out a first investigation of the pixel centroid properties of an H4RG-10 detector array from the {\slshape Roman Space Telescope} flight lot. We developed a method to estimate pixel offsets from speckle patterns, and tested it in simulations, including simulations using the Fresnel diffraction integral that produce the correct variation of the fringe spacing. We then applied the pixel offset reconstruction to the real speckle images, and used an artificial injection technique to estimate the response of the reconstruction to pixel offsets. Although the reconstructions are dominated by noise, correlations of the pixel offset maps obtained from different speckle realizations allow us to determine how much of the pixel offset signal is common across realizations (and thus potentially a real, permanent feature of the detector array). We find that this is 0.0107 pixels RMS of $x$-displacement from a regular grid in a $64\times 64$ pixel patch, if we exclude the (poorly reconstructed) horizontal offset in each row. This residual is reduced to 0.0097 pixels RMS if we use a quadratic fit across each patch. We interpret this result as an upper limit on the physical centroid offsets of the pixels, since any sources of systematic error in the reconstruction (e.g., related to noise correlations or cross-talk) would also contribute.

This result is both encouraging, and suggests that further work to characterize the pixel centroids on {\slshape Roman} detectors is warranted. For example, the formal statistical uncertainty on the position of a 20th magnitude (AB) star in a 140 s exposure in $H$ band (typical of the {\slshape Roman} high latitude survey) is 0.0047 pix RMS using the \citet{2012arXiv1204.5151H} Exposure Time Calculator. The observed pixel offsets at the 0.01 pix RMS level are thus expected to be important for the more demanding {\slshape Roman} science applications. The PSF recovery simulations in Section~\ref{sec:psf}, while simplified, suggest that the $x$-centroid offsets will be a significant contribution to the PSF determination error with {\slshape Roman}. On the other hand, since the H4RG-10s are new and the physical pixel size is smaller than previous generations of NIR detectors (10 $\mu$m instead of 18 $\mu$m), this analysis could have found much larger offsets -- for example, we could have found an offset analogous to the 0.033 pix stitching pattern on the {\slshape Hubble Space Telescope} WFPC2 detectors \citep[e.g.][]{1995ApOpt..34.6672S, 1999PASP..111.1095A} had it been present here and in the $x$-direction (the axis we measured). The fact that such offsets were not found is encouraging.

There are several important limitations to the analysis in this paper. First, it is only a statistical measurement -- an RMS cannot tell us, for example, whether all pixels are affected at the 0.01 pix level or 1/9 of the pixels are affected at the 0.03 pix level. Second, because of the orientation of the apertures used, we have measured only the $x$-offsets of the pixels. The $y$-offsets are of interest, and to the extent that there are apparent pixel shifts due to settling effects, the vertical and horizontal directions would be expected to behave differently due to the readout architecture (see \citealt{2020PASP..132g4504F} for discussion of one such effect in the {\slshape Roman} detectors). The test was done with laboratory electronics and it is possible that in the flight configuration the cross-talk contribution to the apparent pixel centroids would be different. Finally, we have examined one detector array thus far (SCA21536), and it is possible that pixel centroid offsets, like some other properties of the NIR arrays, may vary significantly from one chip to another.

The speckle method developed here has both drawbacks and advantages. The stochastic nature of the scene (where the power spectrum rather than the individual realization is known) limits the precision achievable with a given number of exposures: one is limited by the realization noise ($\Delta I/I\sim 1$) rather than the Poisson noise ($\Delta I/I\sim N_{\rm photon}^{-1/2}$). While one might hope to develop more advanced reconstruction techniques (as was the case for CMB lensing, e.g., \citealt{2003PhRvD..68h3002H, 2020PhRvL.124m1301A, 2021ApJ...922..259M}), a deterministic scene should enable far better constraints. Furthermore, the reconstruction method proved to be very computationally expensive: a typical reconstruction run on the full $4096\times 4096$ detector array takes $\sim 3$ days using 64 cores (2.4 GHz Intel Xeon processors), dominated by linear algebra operations that are already highly optimized. For these reasons, we decided to proceed with projection of a deterministic pattern for the next stage of characterization.

On the other hand, the speckle reconstruction approach has some advantages. The most important for this project was that the data were already available. But it is also possible that, depending on the needs of future projects and the technical constraints for characterization tests, speckle projection may be an attractive option. The speckle pattern does not require any precise alignment or knowledge of the positions of any of the elements (for example, a defocused speckle pattern is another realization of the speckle pattern), although it does require stability over the duration of the measurement (here $\sim 8$ s for the ``half exposure''). The degeneracy where overall row offsets cannot be measured is specific to the double slit aperture geometry, and other aperture shapes that we have tried to simulate (e.g., grids of a few holes) would not be subject to this limitation. The speckle technique may therefore remain an attractive option for statistical characterization of pixel centroid offsets.

\section*{acknowledgements}

We wish to thank Bernie Rauscher and John MacKenty for useful discussions during the development of this project; and Greg Mosby and Ed Wollack for feedback on the draft.

This paper is based on data acquired at the Detector Characterization Laboratory at NASA Goddard Space Flight Center.

This work used results from the {\sc Solid-waffle} detector characterization tools; we thank Ami Choi, Jenna Freudenburg, Jahmour Givans, and Anna Porredon for their contributions to that package.

This work made extensive use of the Pitzer cluster at the Ohio Supercomputer Center \citep{Pitzer2018}.

During the preparation of this work, C.M.H. was supported by NASA award 15-WFIRST15-0008, Simons Foundation award 60052667, and the David \& Lucile Packard Foundation.

\software{astropy \citep{2013A&A...558A..33A,2018AJ....156..123A}, numpy \citep{numpy}, ds9 \citep{2003ASPC..295..489J}, matplotlib \citep{Hunter:2007}
          }


\bibliography{speckle}{}
\bibliographystyle{aasjournal}

\end{document}